\documentclass[11pt,a4paper]{article}
\usepackage{amsmath,amsfonts,amssymb,amsthm,amstext,amscd,array,bbold}
\usepackage[all]{xy}
\DeclareMathOperator{\AdS}{AdS}

\DeclareMathOperator{\NW}{NW}

\let\S\Sphere
\newcommand{\NO}{\mbox{$\substack{\circ\\\circ}$}}      %% Normal
\newcommand{\ee}[1]{{\rm e}^{#1}}
\newcommand{\ii}{{\rm i}}

\newcommand{\mbf}[1]{{\boldsymbol {#1} }}
\def\ii{{\,{\rm i}\,}}
\def\dd{{\rm d}}
\def\DD{{\rm D}}
\def\CC{{\rm C}}
\def\UU{{\rm U}}
\def\BB{{\rm B}}
\def\SU{{\rm SU}}
\def\Diff{{\rm Diff}}

\def\Vect{{\rm Vect}}

\def\mx{{\mbf x}}
\def\my{{\mbf y}}

\def\mdell{{\partial}}

\def\mfnw{{\mathfrak{nw}}}

\def\mfg{{\mathfrak g}}
\def\mfs{{\mathfrak s}}
\def\mfu{{\mathfrak u}}
\def\mff{{\tilde{\mathfrak f}}}
\def\mfff{{\mathfrak f}}

\def\mcD{{\mathcal D}}

\def\mcF{{\mathcal F}}
\def\mcB{{\mathcal B}}
\def\mcU{{\mathcal U}}
\def\mcQ{{\mathcal Q}}
\def\mcI{{\mathcal I}}

\newcommand{\eq}{\begin{equation}}
\newcommand{\eqend}{\end{equation}}
\newcommand{\eqa}{\begin{eqnarray}}
\newcommand{\nonueqa}{\begin{eqnarray*}}
\newcommand{\eqaend}{\end{eqnarray}}
\newcommand{\nonueqaend}{\end{eqnarray*}}

\newcommand{\bma}[1]{\begin{array}{#1}}
\newcommand{\ema}{\end{array}}
\newcommand{\bc}{\begin{center}}
\newcommand{\ec}{\end{center}}

\renewcommand{\thefootnote}{\fnsymbol{footnote}}
\newcommand{\newsection}{\setcounter{equation}{0}\section}

\def\appendix#1{\addtocounter{section}{1}\setcounter{equation}{0}
\renewcommand{\thesection}{\Alph{section}}
\section*{Appendix \thesection\protect\indent \parbox[t]{11.715cm} {#1}}
\addcontentsline{toc}{section}{Appendix \thesection\ \ \ #1} }

\newcommand{\complex}{{\mathbb C}} %% complex numbers
\newcommand{\zed}{{\mathbb Z}} %% integers
\newcommand{\nat}{{\mathbb N}} %% naturals
\newcommand{\real}{{\mathbb R}} %% real numbers

\newcommand{\mat}{{\mathbb M}} %% matrix algebra
\newcommand{\id}{{1\!\!1}} %% identity matrix
\def\alg{{\mathcal A}}
\def\palg{{\mathcal P}}
\def\hil{{\mathcal H}}

\newif\ifold             \oldtrue

\newcommand{\tr}[1]{\:{\rm tr}\,#1}
\newcommand{\Tr}[1]{\:{\rm Tr}\,#1}

\def\e{{\,\rm e}\,}

\def\be{\begin{equation}}
\def\ee{\end{equation}}
\def\bea{\begin{eqnarray}}
\def\eea{\end{eqnarray}}
\def\bd{\begin{displaymath}}
\def\ed{\end{displaymath}}

\newcommand{\beq}{\begin{eqnarray}}
\newcommand{\eeq}{\end{eqnarray}}

\makeatletter
\newdimen\normalarrayskip              % skip between lines
\newdimen\minarrayskip                 % minimal skip between lines
\normalarrayskip\baselineskip
\minarrayskip\jot
\newif\ifold             \oldtrue            
\def\arraymode{\ifold\relax\else\displaystyle\fi} % mode of array entries
\def\@arrayskip{\ifold\baselineskip\z@\lineskip\z@
     \else
     \baselineskip\minarrayskip\lineskip2\minarrayskip\fi}
\def\@arrayclassz{\ifcase \@lastchclass \@acolampacol \or
\@ampacol \or \or \or \@addamp \or
   \@acolampacol \or \@firstampfalse \@acol \fi
\edef\@preamble{\@preamble
  \ifcase \@chnum
     \hfil$\relax\arraymode\@sharp$\hfil
     \or $\relax\arraymode\@sharp$\hfil
     \or \hfil$\relax\arraymode\@sharp$\fi}}
\def\@array[#1]#2{\setbox\@arstrutbox=\hbox{\vrule
     height\arraystretch \ht\strutbox
     depth\arraystretch \dp\strutbox
     width\z@}\@mkpream{#2}\edef\@preamble{\halign \noexpand\@halignto
\bgroup \tabskip\z@ \@arstrut \@preamble \tabskip\z@ \cr}%
\let\@startpbox\@@startpbox \let\@endpbox\@@endpbox
  \if #1t\vtop \else \if#1b\vbox \else \vcenter \fi\fi
  \bgroup \let\par\relax
  \let\@sharp##\let\protect\relax
  \@arrayskip\@preamble}
\makeatother

\begin{document}
\begin{titlepage}
\begin{flushright}

\baselineskip=12pt

HWM--06--29\\
EMPG--06--05\\
hep--th/0606233\\
\hfill{ }\\
June 2006
\end{flushright}

\begin{center}

\vspace{2cm}

\baselineskip=24pt

{\Large\bf Symmetry, Gravity and Noncommutativity\footnote{Invited
  review article to be published in {\sl Classical and Quantum
    Gravity}.}}

\baselineskip=14pt

\vspace{1cm}

{\bf Richard J. Szabo}
\\[4mm]
{\it Department of Mathematics}\\ and \\ {\it Maxwell Institute for
  Mathematical Sciences\\ Heriot-Watt University\\ Colin Maclaurin
  Building, Riccarton, Edinburgh EH14 4AS, U.K.}
\\ {\tt R.J.Szabo@ma.hw.ac.uk}
\\[40mm]

\end{center}

\begin{abstract}

\baselineskip=12pt

We review some aspects of the implementation of spacetime symmetries
in noncommutative field theories, emphasizing their origin in string
theory and how they may be used to construct theories of
gravitation. The geometry of canonical noncommutative gauge
transformations is analysed in detail and it is shown how
noncommutative Yang-Mills theory can be related to a gravity
theory. The construction of twisted spacetime symmetries and their
role in constructing a noncommutative extension of general relativity is
described. We also analyse certain generic features of noncommutative
gauge theories on D-branes in curved spaces, treating several explicit
examples of superstring backgrounds.

\end{abstract}

\end{titlepage}
\setcounter{page}{2}

\newpage

{\baselineskip=12pt
\tableofcontents}

\newpage

\renewcommand{\thefootnote}{\arabic{footnote}} \setcounter{footnote}{0}

\newsection{Introduction\label{Intro}}

General relativity is a dynamical system whose symmetry group contains
general diffeomorphisms of spacetime. The dynamical variable is
spacetime itself equiped with appropriate tensor fields such as a
metric. Upon quantization the classical dynamical variables become
noncommuting operators. This has led to the belief that the classical
differentiable manifold structure of spacetime at the Planck scale
should be replaced by some sort of noncommutative structure. In this
context a proper understanding of quantum gravity requires taking
quantum field theory beyond a framework based on
locality. Noncommutative geometry provides a precise and rigorous
formalism for investigating conceptual problems related to these and
other issues.

The arguments that spacetime noncommutativity appears to be a general
feature of any quantum theory of gravity is most apparent in string
theory, which gives explicit dynamical realizations of the required
non-local smearing out of spacetime coordinates
(see~\cite{Szrev1,Szrev2} and references therein). It describes the
appropriate modification of classical general relativity, and hence of
spacetime symmetries, at short-distance scales. There are several
hints that general covariance emerges in this framework from an
extended {\it gauge} symmetry group. This can already be seen at the
level of closed string dynamics~\cite{LSz1,LLSz1}. The extended
symmetry arises from the low-energy limit of a closed string vertex
operator algebra as a consequence of the holomorphic and
antiholomorphic mixing between closed string states. The
diffeomorphism group of the spacetime acts on the vertex operator
algebra by inner automorphisms, and thereby determines a gauge
symmetry of the low-energy effective field theory. The precise form of
this noncommutative field theory can be described by embedding
D-branes into the background string spacetime, equiped with
appropriate supergravity fields. The low-energy dynamics on the brane
worldvolumes is then governed by a noncommutative deformation of
Yang-Mills gauge theory~\cite{SW1}. D-branes can thus probe Planckian
distances in spacetime where their worldvolume field theories are
drastically altered by quantum gravitational fluctuations.

General quantum mechanical arguments indicate that it is not possible
to measure a classical background spacetime at the Planck scale, due to
the effects of gravitational backreaction~\cite{Szrev2}. It is
therefore tempting to incorporate the dynamical features of spacetime
at a deeper kinematical level using the standard techniques of
noncommutative field theory~\cite{DNrev,Szrev1}. The search for
consistent noncommutative deformations of Einstein gravity has been a
subject of interest for a considerable amount of time. An incomplete
list of references
is~\cite{CFF1,MadoreMourad1,Castellani1,ConnesSpectral1,
  JRam1,Moffat1,Chamdef1,Vacaru1,NishRaj1,Cac1,CZ1,
  Garcia1,Vass1,Madorerecent1}. Particularly noteworthy in this regard
are the gravity theories built on fuzzy
spaces~\cite{Madore2,Nairfuzzy1,Nairfuzzy2,Valt1,Nairfuzzy3,KSfuzzy1},
wherein the noncommutative deformation retains all isometries of the
original classical spacetime. The crucial issues involved in the
construction of any noncommutative theory of gravity is to seek some
guiding dynamical principle for the deformation of general relativity,
and to consistently implement the concept of a general coordinate
transformation in the noncommutative setting.

This article is devoted to an overview of some of these realizations
of gravity in the framework of noncommutative field theory. Our review
is not exhaustive. In particular, we focus only on those features
which emerge from some underlying dynamics, such as the noncommutative
field theories which naturally arise on D-branes in non-trivial string
backgrounds. Roughly half of the paper deals with the simplest case of
flat Moyal spaces. The relevant formalism is briefly reviewed in
Section~\ref{NCFTrev}. In Section~\ref{SymCNGT} we go over some old
material indicating that gravitation is naturally contained in the
gauge-invariant dynamics of noncommutative Yang-Mills theory on flat
space. While some of this material is already reviewed
in~\cite{Szrev1}, we revisit the subject with more of an emphasis on
the manner in which the constructions are reminescent of general
relativity and with some updates on the current points of view. This
analysis is useful for comparison with some of the later sections,
which deal with more current affairs. In
particular, we show how noncommutative Yang-Mills theory naturally
induces a gauge theory of gravitation along the lines described
in~\cite{LangSz1,Madorerecent1,Vacaru1}.

In Section~\ref{TwistSym} we then start turning our attention to some
newer developments, beginning with issues surrounding the breaking of
Lorentz invariance in canonical noncommutative field theories, which
are important for aspects concerning causality and unitarity. The
twist deformation of Poincar\'e spacetime transformations gives
noncommutative field theories a precise meaning of relativistic
invariance. Moreover, the twist procedure naturally extends to give a
deformed Hopf algebra of diffeomorphisms of spacetime in such a way
that the noncommutativity of spacetime is the same in any observer
frame of reference. This allows one to construct a noncommutative
deformation of Einstein's gravity in the standard way. In this
approach general covariance arises as a quantum group symmetry from
the twist deformation of the spacetime symmetries.

In any dynamical theory of gravity, the restriction to flat spacetime
is not natural, and one must eventually discuss more general curved
spacetime manifolds. This is dealt with in Section~\ref{NCGTCurved},
where we analyse in some detail the construction of noncommutative
gauge theories on D-branes in curved backgrounds, and the
implementation of spacetime symmetries in these theories. In
Section~\ref{SuperBacks} we describe three specific superstring
backgrounds as concrete illustrations of the general formalism on
curved noncommutative spaces.

\newsection{Canonical Noncommutative Field Theory\label{NCFTrev}}

In this section we will briefly review the construction of field
theories on Moyal-type (or canonical) noncommutative spaces, mainly to
set up notation. We will do so by emphasizing the two ``dual'' ways of
describing these models, in the sense that there is a one-to-one mapping
between the two descriptions. This point of view will then be
generalized later on to more complicated noncommutative spaces. More
detailed treatments of the material of this section, with exhaustive
lists of references, can be found in~\cite{DNrev,KSrev,Szrev1}.

\subsection{Moyal Product\label{MoyalProd}}

Consider flat euclidean spacetime $\real^D$. Deform the algebra
$\CC^\infty(\real^D)$ of fields on this space by replacing the usual
commutative pointwise multiplication of smooth functions
$f,g:\real^D\to\complex$ by the non-local Moyal star-product, which
may be defined as the formal asymptotic expansion
\beq
(f\star g)(x)=f(x)\,g(x)+\sum_{n=1}^\infty\,\left(\frac\ii2\right)^n\,
\frac1{n!}\,\theta^{i_1j_1}\cdots\theta^{i_nj_n}\,\partial_{i_1}
\cdots\partial_{i_n}f(x)\,\partial_{j_1}\cdots\partial_{j_n}
g(x) \ ,
\label{Moyalprodasymp}\eeq
where $\theta=(\theta^{ij})$ is a constant skew-symmetric $D\times D$
matrix and $\partial_i:=\partial/\partial x^i$ in local coordinates
$x=(x^i)\in\real^D$. Then
$\alg_\theta=\alg_\theta(\real^D):=(\CC^\infty(\real^D)\,,\,\star )$ is an
associative, noncommutative algebra.

The expansion (\ref{Moyalprodasymp}) originates from the
representation of the Moyal product as a {\it twist} deformation of
the ordinary product of functions. Let
\beq
\mu_0\,:\,\alg_0\otimes\alg_0~\longrightarrow~
\alg_0 \ , \quad f\otimes g~\longmapsto~f\,g
\label{mu0commdef}\eeq
be the commutative pointwise product homomorphism on the algebra of
functions $\CC^\infty(\real^D)$. The invertible ``twist'' element
\beq
\mcF_\theta=\exp\left(-\mbox{$\frac\ii2$}\,\theta^{ij}\,\partial_i\otimes
\partial_j\right)
\label{Moyaltwistdef}\eeq
acts on the tensor product $\alg_0\otimes\alg_0$ and belongs to
$U(\real^D)\otimes U(\real^D)$, where $U(\real^D)$ is the universal
enveloping algebra of the translational symmetry algebra of
$\real^D$. Then the star-product (\ref{Moyalprodasymp}) may be
equivalently written as
\beq
f\star g~=~\mu_\theta(f\otimes g)~:=~\mu_0\circ\mcF^{-1}_\theta(f\otimes g)
\label{Moyalprodtwistdef}\eeq
in terms of the noncommutative product map
$\mu_\theta:\alg_\theta\otimes\alg_\theta\to\alg_\theta$. This point of
view will be exploited in Section~\ref{TwistSym}.

The Moyal bracket of two functions is defined to be
\beq
[f,g]_\star ~:=~f\star g-g\star f~=~\ii\{f,g\}_\theta+O\left(\partial^3f\,,\,
\partial^3g\right) \ ,
\label{Moyalbracketdef}\eeq
where
\beq
\{f,g\}_\theta=\theta^{ij}\,\partial_if\,\partial_jg
\label{MoyalPoisson}\eeq
is the Poisson bracket associated to the skew-symmetric form
$\theta$ which defines a constant Poisson structure on $\real^D$. The
Moyal bracket $[-,-]_\star $ makes $\CC^\infty(\real^D)$ into a Lie algebra
which we denote by $\mfu(\alg_\theta)$. It
follows from these definitions that the coordinate generators of
$\alg_\theta$ are noncommuting with the Heisenberg algebra relations
\beq
\left[x^i\,,\,x^j\right]_\star =\ii\theta^{ij} \ .
\label{Heisenalg}\eeq
Moreover, when $\theta$ is nondegenerate (this requires an even
spacetime dimension $D$) translations act as inner derivations of the
noncommutative algebra owing to the identity
\beq
\left[x^i\,,\,f\right]_\star =\ii\theta^{ij}\,\partial_jf \ .
\label{Moyalinnderiv}\eeq

\subsection{Weyl Representation\label{WeylRep}}

Consider the noncommutative space $\real_\theta^D$ defined by
hermitean coordinate generators $\hat x^i$ obeying the canonical
Heisenberg commutation relations $[\hat x^i,\hat
x^j]=\ii\theta^{ij}$. We will now use {\it Weyl quantization} to
systematically associate to any field on $\real^D$ an operator in the
noncommutative algebra generated by the operators $\hat x^i$. Given a
function $f(x)$ on $\real^D$ with Fourier transform $\tilde f(k)$, we
introduce its {\it Weyl symbol} by
\beq
\hat f=\int\frac{\dd^Dk}{(2\pi)^D}~\tilde f(k)~\e^{\ii k\cdot\hat x} \
,
\label{MoyalWeylSymbdef}\eeq
where the symmetric (or Weyl) ordering prescription has been chosen.

Let $f,f_1,\dots,f_n,g\in\CC^\infty(\real^D)$. Let $\Tr$ be a suitably
normalized cyclic trace on the algebra $\real_\theta^D$ of Weyl
operators, for instance a trace over states of a separable Hilbert
space $\hil$ on which $\real_\theta^D$ is represented faithfully by
linear operators. Then one has the following fundamental properties
of the Weyl representation:
\begin{enumerate}
\item Wigner transform: $\displaystyle{
f(x)=\int\frac{\dd^Dk}{(2\pi)^D}~\e^{-\ii k\cdot x}~\Tr\bigl(
\hat f~\e^{\ii k\cdot\hat x}\bigr) \ . }$
\item Algebra isomorphism $\real_\theta^D\cong\alg_\theta(\real^D)$:
$\hat f\,\hat g=\widehat{f\star g} \ . $
\item Integration over noncommutative coordinates $\hat x^i$:
  $\displaystyle{
\Tr\big(\hat f\,\big)=\int\dd^Dx~f(x) \ . }$
\item Cyclicity: $\displaystyle{
\Tr\big(\hat f_1\cdots\hat f_n\big)=\int\dd^Dx~\big(f_1\star \cdots\star f_n
\big)(x) \ . }$
\item $\displaystyle{\int\dd^Dx~(f\star g)(x)=\int\dd^Dx~f(x)\,g(x)} \ . $
\end{enumerate}
The last two properties follow for Schwartz functions on $\real^D$ via
integration by parts.

In addition to the integral defined in Property~3 above, it is also
possible to introduce derivatives in the Weyl representation by
exploiting the translational symmetry of the noncommutative algebra
$\alg_\theta$. Define automorphisms
$\hat\partial_i:\real_\theta^D\to\real_\theta^D$ by
\beq
\bigl[\hat\partial_i\,,\,\hat x^j\bigr]~=~\delta_i{}^j \qquad
\mbox{and} \qquad \bigl[\hat\partial_i\,,\,\hat\partial_j\bigr]~=~0 \
.
\label{Moyalderivdef}\eeq
One then has a covariance property with the Weyl transform
(\ref{MoyalWeylSymbdef}) given by
\beq
\bigl[\hat\partial_i\,,\,\hat f\,\bigr]=\widehat{\partial_if} \ .
\label{covWeylsymb}\eeq
With these ingredients one can now construct and analyse field
theories on the Moyal noncommutative space. We will do this in the
next section by presenting one of our main models of interest in this
paper.

\newsection{Canonical Noncommutative Gauge Symmetries\label{SymCNGT}}

The purpose of this section is to describe to what extent
noncommutative gauge transformations can be interpreted as {\it
  spacetime} symmetries of gauge theories on Moyal noncommutative
spaces. We will describe in detail both the algebraic and geometric
structure of the noncommutative gauge group. We will then
discuss a manner in which these models can serve as gauge theories of
gravitation. More details of noncommutative Yang-Mills theory in the
context of this section can be found in the
reviews~\cite{DNrev,Szrev1}.

\subsection{Star-Gauge Symmetry \label{Defs}}

Let $A=A_i(x)~\dd x^i$ be a ${\rm U}(N)$ gauge field. The action for
noncommutative Yang-Mills theory of $A$ is given by
\beq
S_{\rm NCYM}~:=~-\frac1{4g^2}\,\Tr\otimes\tr\bigl(
\hat F_{ij}^2\bigr)~=~-\frac1{4g^2}\,\int\dd^Dx~
\tr\bigl(F_{ij}(x)^2\bigr) \ ,
\label{SNCYMdef}\eeq
where $g$ is the Yang-Mills coupling constant, $\Tr$ is the operator
trace introduced in Section~\ref{WeylRep}, $\tr$ is the trace over
colour indices, and
\bea
F_{ij}&=& \partial_iA_j-\partial_jA_i-\ii[A_i,A_j]_\star \nonumber\\
&=&\partial_iA_j-\partial_jA_i-\ii[A_i,A_j]+\mbox{$\frac12$}\,
\theta^{kl}\,\bigl(\partial_kA_i\,\partial_lA_j-\partial_kA_j\,
\partial_lA_i\bigr)+O\left(\theta^3\right)
\label{Fijdef}\eeq
is the noncommutative field strength tensor. Note that the action
(\ref{SNCYMdef}) contains an intricate mixing of colour and spacetime
degrees of freedom, in that the spacetime trace (integral) $\Tr$
cannot be separated from the internal ${\rm U}(N)$ trace $\tr$. This
mixing will play a prominent role in this section.

The action (\ref{SNCYMdef}) describes the low-energy effective field
theory for open strings ending on $N$ D-branes in a constant background
$B$-field in the Seiberg-Witten decoupling limit~\cite{SW1}. One of the present
goals is to derive closed string, i.e. gravitational, degrees of freedom
from these open string gauge theories. The action is invariant under
the local {\it star-gauge transformations}
\beq
A_i~\longmapsto~g\star A_i\star g^\dag-\ii g\star \partial_i g^\dag \ ,
\label{stargaugetransf}\eeq
where $g\in\CC^\infty(\real^D,\mat_N)$ is a star-unitary field
\beq
g\star g^\dag~=~g^\dag\star g~=~\id_N \qquad \mbox{equivalently} \qquad
\hat g\,\hat g^\dag~=~\hat g^\dag\,\hat g~=~\id \ .
\label{starunitary}\eeq
The infinitesimal form of the local noncommutative gauge
transformation (\ref{stargaugetransf}) is given by $A_i\mapsto
A_i+\delta^\star _\ell A_i$ with
\beq
\delta^\star _\ell A_i=\partial_i\ell+\ii[\ell,A_i]_\star 
\label{deltaAidef}\eeq
for $\ell\in\mfu(\alg_\theta)$, and the properties of the Moyal
product imply that the linear map $\ell\mapsto\delta^\star _\ell$ is a
representation of the Lie algebra $\mfu(\alg_\theta)$,
\beq
\big[\delta^\star _\ell\,,\,\delta^\star _{\ell'}
\big]=\delta^\star _{[\ell,\ell']_\star }
\label{deltaellLiealgrep}\eeq
for all $\ell,\ell'\in\mfu(\alg_\theta)$. Observables of
noncommutative gauge theory, respecting the symmetry
(\ref{stargaugetransf}), are provided by open and closed Wilson line
operators. They will not be dealt with at length in this article.

\subsection{Geometry of Star-Gauge
  Transformations\label{StarGaugeGeom}}

We will now begin to identify the geometrical implications of the huge
noncommutative gauge symmetry. The goal is to capture the manner in
which noncommutative Yang-Mills theory can serve as a gauge theory of
gravity. A preliminary indication that this
may be possible is by realizing that spacetime translations can be
implemented through star-gauge transformations.

Assume that the dimension $D=2d$ is even and that the tensor
$\theta$ is of maximal rank $d$. There is no loss of generality in
only analysing the simplest case of $\UU(1)$ gauge theory. Consider
the plane wave
\beq
g_\ell(x)=\e^{-\ii\ell^i\,B_{ij}\,x^j}
\label{Uellxdef}\eeq
where $\ell=(\ell^i)\in\real^D$ is a constant vector and
$B_{ij}=(\theta^{-1})_{ij}$ are the components of the constant
background supergravity $B$-field in the topological limit where the
bulk closed string metric completely decouples~\cite{SW1,Szrev2}. It
is star-unitary
\beq
g_\ell(x)\star g_\ell(x)^\dag~=~g_\ell(x)^\dag\star g_\ell(x)~=~1 \ ,
\label{Uellxstarunitary}\eeq
and using (\ref{Moyalinnderiv}) along with the
Baker-Campbell-Hausdorff formula shows that it generates translations
of scalar fields $f$ through star-conjugation
\beq
g_\ell(x)\star f(x)\star g_\ell(x)^\dag=f(x+\ell) \ .
\label{Uellxtranslgen}\eeq
The corresponding gauge transformation (\ref{stargaugetransf}) reads
\beq
A_i(x)~\longmapsto~A_i(x+\ell)-B_{ij}\,\ell^j \ .
\label{gaugetransftransl}\eeq
It follows that up to a global symmetry transformation of the field
theory, under which the field strength tensor $F_{ij}(x)$ is
invariant, spacetime translations are equivalent to gauge
transformations. Noncommutative gauge theories are thus ``toy models''
of general relativity~\cite{Grosstoy}. To make this more precise one
needs to gauge the global translational symmetry and repeat the
constructions with generic non-constant functions $\ell^i=\ell^i(x)$
on $\real^D$. These functions correspond to more general spacetime
transformations which we may wish to compare with diffeomorphisms of
$\real^D$. We will describe how to do this in
Section~\ref{Teleparallel} below.

Superficially, such a construction does not appear to be possible for
the following reason. Consider an infinitesimal unitary transformation
of a scalar field $f$ by a function $\ell(x)$ on $\real^D$ given by
$\delta^\star _\ell f:=\ii[f,\ell]_\star $. By (\ref{Moyalbracketdef})
it coincides at leading order in the limit $\theta\to0$ (or
equivalently for slowly-varying fields) with the Poisson bracket
$\{\ell,f\}_\theta$. It follows that the gauge group of noncommutative
Yang-Mills theory in this limit coincides with the group of canonical
transformations preserving the Poisson structure $\theta$ on
$\real^D$, i.e. with the Poisson diffeomorphism group
$\Diff_\theta(\real^D)$. The higher-derivative terms in
(\ref{Moyalbracketdef}) modify this interpretation in a way that we
describe explicitly in Section~\ref{NCGaugeGp} below.

The crucial issue is the closure of the set of gauge
functions to a group. For example, the set of linear functions of the
form (\ref{Uellxdef}) close a group with respect to the star-product,
since a simple computation using the Baker-Campbell-Hausdorff formula
shows that
\beq
g_{\ell_1}\star 
g_{\ell_2}=\e^{-\frac\ii2\,\ell_1^i\,B_{ij}\,
\ell_2^j}~g_{\ell_1+\ell_2} \ .
\label{gellclosure}\eeq
More generally, arbitrary linear transformations $x\mapsto Lx$,
$L\in{\rm GL}(D,\real)$ are implementable as gauge symmetries and the
corresponding generators close a group. In fact, the most general
gauge functions which close a group correspond to bilinear
forms $x\cdot Qx+\xi\cdot x$ with $Q\in{\rm GL}(D,\real)$ symmetric and
$\xi\in\real^D$~\cite{LSZ1}. However, the only spacetime symmetries
which preserve the star-product of two fields, and hence define
automorphisms of the algebra $\alg_\theta$, are linear affine
transformations $x\mapsto Lx+\ell$. These transformations act on the
space of antisymmetric matrices $\theta$ as congruence $\theta\mapsto
L\,\theta\,L^\top$. The Moyal product is thus fully covariant under
linear affine transformations~\cite{G-BLR-RV1}, reflected in the
algebra isomorphisms
$\alg_\theta^{~}\cong\alg_{L\,\theta\,L^\top}$. Demanding that the
noncommutative Yang-Mills action be invariant further restricts to
transformations of unit jacobian. Thus only the subgroup ${\rm
  SL}(D,\real)\ltimes\real^D\subset\Diff_\theta(\real^D)$ of
unimodular linear affine maps appear to be gauge symmetries. In the
next section we will see how to overcome these and other restrictions
of spacetime symmetries generally in noncommutative field theory.

\subsection{Automorphisms\label{Automor}}

The mixing between spacetime and internal gauge symmetries can be best
understood in an abstract setting by examining the {\it automorphism
  group} ${\rm Aut}_N(\alg_\theta):={\rm
  Aut}(\alg_\theta\otimes\mat_N)$ of the algebra
$\alg_\theta\otimes\mat_N$ of $N\times N$ matrix-valued fields on
$\real^D$ equiped with the star-product. The group ${\rm
  Aut}_N(\alg_\theta)$ has a natural normal subgroup ${\rm
  Inn}_N(\alg_\theta)$ consisting of {\it inner automorphisms}
\beq
f~\longmapsto~g\star f\star g^\dag\qquad \mbox{with} \qquad f\in\alg_\theta
\otimes\mat_N \quad \mbox{and} \quad g\in\UU_N(\alg_\theta) \ ,
\label{innerautos}\eeq
where $\UU_N(\alg_\theta)$ is the unitary group of the algebra
$\alg_\theta\otimes\mat_N$ consisting of those matrix fields $g$ which
obey (\ref{starunitary}). The remaining automorphisms
comprise the group of {\it outer automorphisms} ${\rm
  Out}_N(\alg_\theta)$ such the full automorphism group is the
semi-direct product
\beq
{\rm Aut}_N(\alg_\theta)={\rm Inn}_N(\alg_\theta)\rtimes
{\rm Out}_N(\alg_\theta) \ .
\label{Autsemidirect}\eeq
If the algebra $\real^D_\theta$ is represented faithfully on a
separable Hilbert space $\hil$, then these groups are related to the
group $\UU(\hil)$ of unitary operators on $\hil$.

Consider, for example, the case of $\UU(1)$ gauge theory on the
commutative space $\real^D$, i.e. the automorphisms of the algebra
$\alg_0=\CC^\infty(\real^D)$. We may represent $\alg_0$ on its dense
subspace $\hil={\rm L}^2(\real^D)$ of square-integrable fields by
multiplication
\beq
m_f\,:\,\psi~\longmapsto f\,\psi \qquad \mbox{for} \qquad
f\in\alg_0 \quad \mbox{and} \quad \psi\in\hil \ .
\label{A0L2rep}\eeq
Since $\alg_0$ is commutative, there are no
non-trivial inner automorphisms. On the other hand, outer
automorphisms $\alpha_\phi:\alg_0\to\alg_0$ correspond to smooth
invertible maps $\phi:\real^D\to\real^D$ with
\beq
\alpha_\phi(f)=f\circ\phi^{-1}
\label{alphaphifdef}\eeq
for $f\in\alg_0$. It follows that
\beq
{\rm Inn}(\alg_0)~=~\{\id\} \qquad \mbox{and} \qquad {\rm
  Out}(\alg_0)~=~\Diff\big(\real^D\big) \ .
\label{InnOutA0}\eeq
Corresponding to each outer automorphism we define a unitary operator
$\hat g_\phi$ on $\hil$ by
\beq
\hat g_\phi\psi\big(x\big)=\left|\frac{\partial\phi}{\partial x}
\right|^{1/2}\,\psi\big(\phi^{-1}(x)\big)
\label{hatgphidef}\eeq
such that
\beq
\alpha_\phi(f)=\hat g_\phi\,f\,\hat g_\phi{}^\dag \ .
\label{alphaphifgrep}\eeq
More generally, ${\rm Inn}_N(\alg_0)=\CC^\infty(\real^D\,,\,\UU(N))$
is the usual group of $\UU(N)$ gauge transformations in ordinary
Yang-Mills theory on $\real^D$.

At the other extreme is a finite-dimensional algebra $\mat_N$, for
which {\it all} automorphisms can be represented as rotations by
$N\times N$ unitary matrices and one has
\beq
{\rm Inn}(\mat_N)~=~\UU(N)/\UU(1) \qquad \mbox{and} \qquad
{\rm Out}(\mat_N)~=~\{\id\} \ .
\label{autoMN}\eeq
For the Moyal space, the group of star-gauge transformations ${\rm
  Inn}_N(\alg_\theta)$ realizes a non-trivial mixing between the two
automorphism groups ${\rm Out}(\alg_0)$ and ${\rm Inn}(\mat_N)$ in
(\ref{InnOutA0}) and (\ref{autoMN}). The mixing between spacetime and
matrix degrees of freedom here motivates an interpretation in terms of
matrix models, which we now describe.

\subsection{Matrices\label{Matrices}}

The well-known remarkable feature that noncommutative gauge theory can
be reformulated as a zero-dimensional matrix
model~\cite{NCYMIIB,AMNS1,AMNS2} may be exploited in
the present context to give some insight into the structure of the
group of noncommutative gauge transformations. Consider the
rank one case $N=1$. Introduce the {\it covariant coordinates}
\beq
C_i=B_{ij}\,x^j+A_i
\label{covcoordsdef}\eeq
with the gauge transformations
\beq
C_i~\longmapsto~g\star C_i\star g^\dag \qquad \mbox{and} \qquad
\delta^\star _\ell C_i~=~\ii[\ell,C_i]_\star  \ .
\label{Cigaugetransf}\eeq
Using the inner derivation property (\ref{Moyalinnderiv}), the entire
structure of noncommutative gauge theory can be expressed in terms of
the operators (\ref{covcoordsdef}) in such a way that spacetime
derivatives completely disappear. For example, covariant derivatives
may be rewritten as
\beq
D_if~:=~\partial_if-\ii[A_i,f]_\star ~=~\ii[f,C_i]_\star 
\label{covderivsCi}\eeq
while the field strength tensor (\ref{Fijdef}) can be expressed as
\beq
F_{ij}=-\ii[C_i,C_j]_\star +B_{ij} \ .
\label{FijCi}\eeq

The $C_i$ are elements of the {\it abstract} algebra
$\real^D_\theta$. Passing to the Weyl representation $\hat C_i$, the
action (\ref{SNCYMdef}) becomes
\beq
S_{\rm NCYM}=-\frac1{4g^2}\,\Tr\,\sum_{i\neq j}\,\left(-\ii
\big[\hat C_i\,,\,\hat C_j\big]+B_{ij}
\right)^2 \ .
\label{SNCYMreduced}\eeq
Since $\hat C_i$ are formally space-independent, we have thus found
that noncommutative gauge theory is equivalent to an
infinite-dimensional matrix model~\cite{NCYMIIB}. This is called a {\it twisted
  reduced model}, where the ``twist'' $B_{ij}$ removes an
infinite constant in the rewriting and selects a non-trivial vacuum
for the matrix model (\ref{SNCYMreduced}). It is a large $N$ version of the
IKKT matrix model which describes the nonperturbative dynamics of
Type~IIB superstring theory~\cite{IKKT1}. The spacetime dependence is
hidden in the infinitely-many degrees of freedom of the large $N$
matrices $\hat C_i$. The classical ground state $\hat C_i^{(0)}$ of
(\ref{SNCYMreduced}) generates a Heisenberg algebra $\big[\hat
C_i^{(0)},\hat C_j^{(0)}\big]=-\ii B_{ij}$. Expanding the infinite
matrices $\hat C_i$ around this vacuum enables one to rederive the
noncommutative gauge theory (\ref{SNCYMdef}) from the matrix model
(\ref{SNCYMreduced})~\cite{NCYMIIB}.

While any operator realization of noncommutative gauge theory is
{\it formally} a matrix model, we can actually go further and write
down a {\it finite}-dimensional version which can serve as a regulated
noncommutative quantum field theory beyond perturbation
theory~\cite{AMNS1,AMNS2}. A regulated, $N\times N$ matrix model with
these properties is provided by the {\it twisted Eguchi-Kawai model}
\beq
S_{\rm TEK}=-\frac1{4g^2}\,\sum_{i\neq j}\,Z_{ij}{}^*\,\tr\bigl(
U_i\,U_j\,U_i{}^\dag\,U_j{}^\dag\bigr) \ ,
\label{STEKdef}\eeq
where $U_i\in\UU(N)$, $i=1,\dots,D$ and the twists are given by
\beq
Z_{ij}~=~\e^{2\pi\ii Q_{ij}/N} \qquad \mbox{with} \qquad
Q_{ij}~=~-Q_{ji}\in\zed \ .
\label{TEKtwistsdef}\eeq
The action (\ref{STEKdef}) possesses the gauge symmetry
\beq
U_i~\longmapsto~V\,U_i\,V^\dag \qquad \mbox{with} \qquad
V\in\UU(N) \ .
\label{TEKgaugetransf}\eeq

Let $\epsilon$ be a dimensionful lattice spacing and identify
$U_i=\e^{\ii\epsilon\,\hat C_i}$. Then the action (\ref{STEKdef})
reduces to (\ref{SNCYMreduced}) in the double-scaling continuum limit
$\epsilon\to0$, $N\to\infty$ with
\beq
B_{ij}=\frac{2\pi\,Q_{ij}}{N\,\epsilon^2} \ .
\label{thetaQid}\eeq
Thus the twisted Eguchi-Kawai model is the natural non-perturbative
version of noncommutative Yang-Mills theory. It can be thought of as
originating from the one-plaquette reduction of ordinary Wilson
lattice gauge theory in $D$ dimensions with multivalued gauge fields
and the integers $Q_{ij}$ corresponding to background 't~Hooft
fluxes. This proves that noncommutative gauge theory is equivalent to
a twisted large $N$ reduced model, i.e. the IIB matrix model with
D-brane backgrounds, to {\it all} orders of perturbation theory.

We will use this identification below to give a precise geometric
interpretation to noncommutative gauge transformations. The key
feature is that the gauge fields of the finite-dimensional matrix
model can be expanded in a canonical basis of matrices suitable for
this investigation. The Weyl basis
for the Lie algebra ${\mathfrak{gl}}(N,\complex)$ is given by
\beq
J_k=\prod_{i=1}^D\,(\Gamma_i)^{k_i}\,\prod_{j<i}\,\e^{\pi\ii
k_i\,Q_{ij}\,k_j/N} \ ,
\label{WeylbasisglN}\eeq
where $k=(k_i)\in\zed^D$ are discrete momenta and $\Gamma_i$ are
twist-eating solutions for ${\rm SU}(N)$ which obey the Weyl algebra
in $D$ dimensions
\beq
\Gamma_i\,\Gamma_j=Z_{ij}~\Gamma_j\,\Gamma_i \ .
\label{WeylalgD}\eeq
The matrices (\ref{WeylbasisglN}) obey the orthonormality and
completeness relations
\beq
\frac1N\,\tr\bigl(J_k\,J_q{}^\dag\bigr)~=~\delta_{k,q} \qquad
\mbox{and} \qquad \frac1N\,\sum_k\,(J_k)_{ab}\,(J_k)_{cd}~=~
\delta_{ad}\,\delta_{bc} \ ,
\label{Jkorthocompl}\eeq
where the sum runs over momenta restricted to a Brillouin
zone. They also obey the product rule
\beq
J_k\,J_q=\prod_{i,j=1}^D\,\e^{\pi\ii k_i\,Q_{ij}\,q_j/N}~J_{k+q}
\label{Jkprodrule}\eeq
and thus may be thought of as discrete versions of the plane wave
generators $\e^{\ii k\cdot\hat x}$ of the Weyl representation of
Section~\ref{WeylRep}. In particular, the gauge fields of the twisted
Eguchi-Kawai model (\ref{STEKdef}) may be expanded as
\beq
U_i=\frac1{N^2}\,\sum_k\,U_i(k)~J_k
\label{TEKfieldsexpJk}\eeq
with c-number Fourier coefficients given by
$U_i(k)=N\,\tr(U_i\,J_k{}^\dag)$.

\subsection{Noncommutative Gauge Group\label{NCGaugeGp}}

We can now make our first putative connection between gravitation and
star-gauge symmetries. Represent the algebra $\real_\theta^D$
on a separable Hilbert space $\hil$. We have seen in this section that
there are two natural candidate gauge groups $\UU(\alg_\theta)$ of
$\UU(1)$ noncommutative Yang-Mills theory on $\real^D$. Firstly, from
our discussion in Section~\ref{Automor} above there
is the unitary group $\UU(\hil)$ of the Hilbert space $\hil$. However,
by Kuiper's theorem $\UU(\hil)$ is contractible. In particular, all
of its homotopy groups are trivial. So the group $\UU(\hil)$ doesn't
carry any topology and we lose all of the topological effects, such as
solitons and anomalies, that noncommutative gauge theories are known
to possess. Secondly, from the matrix model formalism of
Section~\ref{Matrices} above there is the infinite unitary group
$\UU(\infty)$. However, $\UU(\infty)$ consists of arbitrarily large
but {\it finite}-dimensional unitary operators. Since
$\UU(\alg_\theta)$ is a group of functions on $\real^D$, it cannot be
generated by finite-dimensional matrices.

Nevertheless, the group $\UU(\infty)$ does have the right
properties that we are looking for. In particular,
$\UU(\infty)\supset\UU(N)$ for all $N$ and it has homotopy
groups determined by Bott periodicity to be
\beq
\pi_n\bigl(\UU(\infty)\bigr)=\left\{\begin{matrix}
~\zed~, & n~\mbox{odd} \\ ~0~, & n~\mbox{even}\end{matrix}\right. \ .
\label{Bottper}\eeq
The key to relating the infinite unitary group $\UU(\infty)$ to
$\UU(\alg_\theta)$ is the continuum limit of the matrix model that we
took in Section~\ref{Matrices} above. In functional analytic terms, it
means that we should {\it complete} $\UU(\infty)$ in the Schatten
$p$-norms on the endomorphism algebra ${\rm End}(\hil)$ for $1\leq
p\leq\infty$. View $\UU(\infty)\subset\UU(\hil)$ as the group of all
finite-rank unitary operators on $\hil$, and define the Schatten norms
\beq
\bigl\|\hat f\,\big\|_p~=~\left(\Tr\big(\hat f^\dag\,\hat f\,\big)^{p/2}\,
\right)^{1/p} \qquad \mbox{for} \qquad 1\leq p<\infty
\label{Schattenpnorm}\eeq
and the operator norm
\beq
\bigl\|\hat f\,\big\|_\infty=\sup_{\langle\psi|\psi\rangle\leq1}\,
\bigl\langle\hat f\psi\,\big|\,\hat f\psi\bigr\rangle^{1/2} \ .
\label{opnormdef}\eeq

Denote the corresponding completions of $\UU(\infty)$ by
$\UU_p(\hil)$. Then there is a sequence of completions of unitary
subgroups of $\UU(\hil)$ given by
\beq
\UU(\infty)~\subset~\UU_1(\hil)~\subset~\UU_2(\hil)~\subset~
\cdots~\subset~\UU_\infty(\hil) \ .
\label{Ucomplseq}\eeq
Writing a generic unitary operator $\hat U$ in the form $\hat
U=\e^{\ii\hat K}$, the operators $\hat K$ corresponding to the sequence
(\ref{Ucomplseq}) are respectively finite-rank, trace-class,
Hilbert-Schmidt, and compact. In particular, the group
$\UU_\infty(\hil)$ consists of unitary operators whose sequence of
eigenvalues approaches~$1$. Under the Weyl-Wigner correspondence
of Section~\ref{WeylRep}, these spaces of operators map
naturally onto ${\rm L}^p$-spaces of functions on
$\real^D$. They respectively give functions $K$ which are integrable
($p=1$), square-integrable ($p=2$), and of rapid fall-off at infinity
in $\real^D$ ($p=\infty$).

In particular, for $p=\infty$ the group consists of unitary operators
which are connected to the identity. In a euclidean path integral
formulation of the quantum gauge theory, the gauge orbit space that
one should integrate over is the quotient of the space of gauge field
configurations on $\real^D$ by the group of gauge transformations
which are connected to the identity. This connectedness property is
thus possessed by the group of compact unitaries
$\UU_\infty(\hil)$. Moreover, in this way we have provided a direct
relationship between the topology of $\UU_\infty(\hil)$ and the
topology of the configuration space of noncommutative gauge
fields~\cite{NP1,Harvey1}. By Palais' theorem, the completion groups in
(\ref{Ucomplseq}) all have the homotopy type of $\UU(\infty)$. We
conclude finally that the noncommutative gauge group is given by
\beq
\UU(\alg_\theta)=\UU_\infty(\hil) \ .
\label{NCgaugegpfinal}\eeq

We now provide a geometrical interpretation of the group
(\ref{NCgaugegpfinal})~\cite{LSZ1}. This is where the matrix model formalism of
Section~\ref{Matrices} above can be put to good use. From the product
rule (\ref{Jkprodrule}) it follows that the generators of the Weyl
basis for $\mathfrak{gl}(N,\complex)$ satisfy the commutation
relations of a trigonometric Lie algebra
\beq
\bigl[J_k\,,\,J_q\bigr]=2\ii\sin\Big(\mbox{$\frac{\pi\ii}N\,
\sum\limits_{i<j}\,k_i\,Q_{ij}\,q_j$}\Big)~J_{k+q} \ .
\label{trigLiealg}\eeq
Take the limit $N\to\infty$ with the momenta $k_i,q_j\ll\sqrt
N$. After an appropriate overall rescaling of the $J_k$, the algebra
(\ref{trigLiealg}) becomes the ${\rm W}_\infty$-algebra
\beq
\bigl[J_k^\infty\,,\,J_q^\infty\bigr]=2\pi\ii k\wedge q~
J_{k+q}^\infty
\label{Winftyalg}\eeq
with $k\wedge q:=k_i\,\theta^{ij}\,q_j$. A detailed, rigorous
description of this large $N$ limit can be found
in~\cite{LLS1,Szrev3}.

This coincides with the Lie algebra of canonical transformations on
$\real^D$ with the constant Poisson structure $\theta$. These are the
diffeomorphisms
\beq
f~\longmapsto~\delta_\phi f~:=~X_\phi(f)~=~\{\phi,f\}_\theta
\label{fcantransfexpl}\eeq
parameterized by scalar fields $\phi\in\CC^\infty(\real^D)$. They are
generated by the Poisson vector fields
\beq
X_\phi=\theta^{ij}\,\partial_i\phi~\mbox{$\frac\partial{\partial
    x^j}$}
\label{Poissonvector}\eeq
which close the Poisson-Lie algebra
\beq
[X_\phi,X_{\phi'}]=X_{\{\phi,\phi'\}_\theta}
\label{PoissonLiealg}\eeq
as a subalgebra of the Lie algebra ${\rm Vect}(\real^D)$ of vector
fields on $\real^D$. Taking $\phi(x)=\phi_k(x)=\e^{2\pi\ii k\cdot x}$,
the Poisson-Lie algebra (\ref{PoissonLiealg}) of the vector fields
$X_k=X_{\phi_k}$ coincides with (\ref{Winftyalg}). As in
Section~\ref{StarGaugeGeom}, fields with high-momentum modes modify
this result.

We conclude that the gauge group (\ref{NCgaugegpfinal})
is a quantum deformation of the Poisson diffeomorphism group
$\Diff_\theta(\real^D)$~\cite{LSZ1}, and in this way we arrive at a
noncommutative {\it unimodular} theory of gravitation (general
relativity based on volume-preserving diffeomorphisms). This point of
view is exploited in~\cite{Calmet1,Calmet2,MS1} to examine the
noncommutative corrections to Einstein's general relativity. This
result has a natural physical interpretation in terms of the
representation of a D-brane as a configuration of infinitely many
lower-dimensional D-branes~\cite{Ishibashi1}. In this case the
$\UU(1)$ gauge theory on the brane induces a $\UU_\infty(\hil)$ gauge
symmetry in the lower-dimensional theory. This can be captured more
quantitatively by coupling gauge theory operators to closed string
states using open Wilson
lines~\cite{Closedcoupl1,Closedcoupl2,Closedcoupl3,Closedcoupl4,Closedcoupl5}.
We will encounter deformed diffeomorphism groups within a more general
framework in the next section.

\subsection{Teleparallelism in Noncommutative Gauge
  Theory\label{Teleparallel}}

Our final point of analysis in this section will be a description of
how the Poisson symmetries inherent in noncommutative Yang-Mills
theory can be extended to more general diffeomorphisms. The idea is to
exploit the mixing of internal and spacetime symmetries in a way which
enables the unambiguous identification of gauge transformations as
general coordinate transformations. Although the gauge group of
noncommutative Yang-Mills theory does not admit a local translational
symmetry corresponding to generic diffeomorphisms of flat space, we
will see that a particular reduction of noncommutative gauge theory
captures the qualitative manner in which noncommutative gauge
transformations realize general covariance. The crux of the
construction is that the Lie algebra
$\mfu(\alg_\theta)=(\CC^\infty(\real^{2d})\,,\,[-,-]_\star )$ of functions
on the space $\real^{2d}$ equiped with the Moyal bracket
(\ref{Moyalbracketdef}) contains the Lie algebra $\Vect(\real^d)$ of
vector fields on a subspace $\real^d\subset\real^{2d}$, where we
identify $\real^{2d}$ with the tangent bundle $T\real^d$. It is then
possible to restrict the noncommutative gauge
fields so as to obtain a local field theory whose symmetry group
incorporates diffeomorphism invariance. Gauge theories which induce
noncommutative gauge theories in lower dimensions are also studied
in~\cite{NR1,NR2}.

We can motivate the ensuing construction by considering the
homogeneous gauge transformation laws (\ref{Cigaugetransf}) obeyed by
the covariant coordinates $C_i$. Given an
arbitrary local vector field $X=X^i(x)~\partial_i$ on $\real^D$, we
introduce a corresponding gauge function
\beq
\ell~=~\ell_X~=~-\ii X^i\,B_{ij}\,x^j \ .
\label{ellXdef}\eeq
The corresponding infinitesimal gauge transformation in
(\ref{Cigaugetransf}) can be computed to leading order in an
asymptotic expansion in $\theta$ with the result
\beq
\delta^\star _{\ell_X}C_i=X(C_i)+B_{kj}\,x^j\,\delta^{lm}\,\delta^{np}\,
\theta_{mp}\,\partial_lX^k\,\partial_nC_i+O(\theta) \ .
\label{deltaellXCi}\eeq
The first term in (\ref{deltaellXCi}) is close to the expected
transformation law for $C_i$ under an infinitesimal diffeomorphism,
except that it treats $C_i$ as a scalar field. As explained above, this
is only consistent for Poisson vector fields $X$ obeying ${\rm
  div}(X)= \partial_iX^i=0$. The second term in (\ref{deltaellXCi}) is
of the same order in $\theta$, and this fact on its own prevents one
from realizing arbitrary diffeomorphisms in terms of star-gauge
transformations. Nevertheless, if one attempts to interpret the first
term in (\ref{deltaellXCi}) as the transformation rule for a flat
space frame field $e_i^j$ defined through the decomposition
$C_i=-x^k\,B_{kj}\,e_i^j$, then the spacetime coordinates themselves
must gauge transform as $\delta^\star _{\ell_X}x^i=X^i(x)$. Unless the vector
field $X$ is parameterized by an element of the Lie algebra
$\complex\oplus\mathfrak{sp}(D)$ as explained in
Section~\ref{StarGaugeGeom} above, such a transformation will map the
Moyal space $\real_\theta^D$ onto a different noncommutative space and
will not be a symmetry of the theory. One may try to find an extended
matrix model with a larger symmetry group than the $\UU(\infty)$ of
Section~\ref{Matrices}~\cite{Supermatrix}. Such an extension is the
essential idea behind the construction which follows.

Define a reduction of noncommutative Yang-Mills theory as
follows~\cite{LangSz1}. Denote the local coordinates of
$\real^{2d}$ by $\xi=(\xi^i)=(x^\mu,y^a)_{\mu,a=1}^d$. The space
$x=(x^\mu)\in\real^d$ is our target spacetime while the $y^a$ can be
interpreted as local coordinates along the fibres of the cotangent
bundle $T^*\real^d$. The noncommutativity parameters are taken to be
of the block form
\beq
\theta=\begin{pmatrix}0 & \theta^{\mu b} \\
\theta^{a\nu} & 0 \end{pmatrix} \ ,
\label{thetablockform}\eeq
and we assume that $(\theta^{\mu b})$ is an invertible $d\times d$
matrix. Having $\theta^{\mu\nu}\neq0$ in (\ref{thetablockform}) would
lead to a gravitational field theory on a noncommutative space, which
will be studied in Section~\ref{TwistDiff}. Consider the linear
subspace $\mfg\subset\mfu(\alg_\theta)$ of smooth functions $\ell$ on
$\real^{2d}$ which are linear in the coordinates $y$,
\beq
\ell(\xi)=\ell_a(x)\,y^a \ .
\label{ellgenreduced}\eeq
The Moyal bracket of any two elements $\ell,\ell'\in\mfg$ is given by 
\beq
[\ell,\ell'\,]_\star (\xi)~=~\bigl([\ell,\ell'\,]_\star \bigr)_a(x)\,y^a \qquad
\mbox{with} \qquad \bigl([\ell,\ell'\,]_\star \bigr)_a~=~\theta^{\mu b}\,
\big(\ell'_b\,\partial_\mu\ell_a-\ell_b\,\partial_\mu\ell'_a\big) \ ,
\label{Moyalbracketred}\eeq
and consequently $(\mfg\,,\,[-,-]_\star )$ is a Lie algebra. Now define the
invertible linear map
\beq
\mfg~\longrightarrow~\Vect\big(\real^d\big) \ , \quad \ell~\longmapsto~
X_\ell=-\theta^{\mu a}\,\ell_a~\mbox{$\frac\partial{\partial x^\mu}$}
\label{linmapVectRd}\eeq
onto the linear space of vector fields on $\real^d$. Then by
(\ref{Moyalbracketred}) it defines a representation of the Lie algebra
$\mfg$,
\beq
[X_\ell,X_{\ell'}]=X_{[\ell,\ell']_\star }
\label{XmfgrepLie}\eeq
for all $\ell,\ell'\in\mfg$, and so $\mfg$ can be identified under the
linear isomorphism (\ref{linmapVectRd}) with the Lie algebra of
diffeomorphisms of $\real^d$ which are connected to the identity.

Define a corresponding trunction of the affine space of
$\UU(1)$ gauge fields $A=A_i(\xi)~\dd\xi^i$ on $\real^{2d}$ by
\beq
A=\omega_{\mu a}(x)\,y^a~\dd x^\mu+\xi_a(x)~\dd y^a \ .
\label{AYMconsred}\eeq
This is the minimal consistent reduction which is closed under the
action of the reduced star-gauge group. It is straightforward to
compute that the gauge transforms (\ref{deltaAidef}) with gauge
functions (\ref{ellgenreduced}) preserve the subspace of gauge fields
of the form (\ref{AYMconsred}), and that the components transform as
\bea
\delta^\star _\ell\omega_{\mu a}&=&\partial_\mu\ell_a+\theta^{\nu b}\,
(\ell_b\,\partial_\nu\omega_{\mu a}-\omega_{\mu b}\,
\partial_\nu\ell_a) \ , \nonumber \\
\delta^\star _\ell \xi_a&=&\ell_a-\theta^{\mu b}\,\ell_b\,\partial_\mu \xi_a
\label{compredtransfs}\eea
for $\ell\in\mfg$. The curvature components (\ref{Fijdef}) of the
gauge field (\ref{AYMconsred}) are also easily computed to be
\bea
F_{\mu\nu}(\xi)&=&\Omega_{\mu\nu a}(x)\,y^a \qquad \mbox{with} \qquad
\Omega_{\mu\nu a}~=~\partial_\mu\omega_{\nu a}-
\partial_\nu\omega_{\mu a}+\theta^{\lambda b}\,(\omega_{\nu b}\,
\partial_\lambda\omega_{\mu a}-\omega_{\mu b}\,\partial_\lambda
\omega_{\nu a}) \ , \nonumber\\
F_{\mu a}&=&\partial_\mu \xi_a-\omega_{\mu a}-\theta^{\nu b}\,
\omega_{\mu b}\,\partial_\nu \xi_a \ , \nonumber\\
F_{ab}&=&0 \ .
\label{curvcompsred}\eea

The truncated fields above are naturally related to the geometry of
spacetime as follows. From the inner
derivation property (\ref{Moyalinnderiv}) and the choice
(\ref{thetablockform}) it follows that the commuting coordinates $y^a$
may be identified with the holonomic derivative generators
$-\theta^{\mu a}\,\partial_\mu$ of the $d$-dimensional translation
group of $\real^d$. Promoting this global symmetry to a local gauge
symmetry as explained in Section~\ref{StarGaugeGeom} is tantamount to
replacing global translations with local translations $x^\mu\mapsto
x^\mu+\ell^\mu(x)$ of the fibre coordinates of the tangent
bundle. This requires the replacement of the derivatives
$\partial_\mu$ with the covariant derivatives
\beq
\nabla_\mu=\partial_\mu+\omega_{\mu a}\,y^a \ ,
\label{nablamudef}\eeq
where $\omega_{\mu a}$ are gauge fields corresponding to the gauging
of the translation group, i.e. to the replacement of $\real^d$ by the
Lie algebra $\mfg$. For any scalar field $f$ one then has
\beq
\nabla_\mu f=e_\mu^\nu\,\partial_\nu f
\label{nablamuf}\eeq
with
\beq
e_\mu^\nu=\delta_\mu^\nu-\theta^{\nu a}\,\omega_{\mu a} \ .
\label{emunudef}\eeq
Using (\ref{linmapVectRd}) one sees that the covariance requirement
\beq
\delta^\star _\ell(\nabla_\mu f)=X_\ell(\nabla_\mu f)
\label{covreqnablamu}\eeq
is equivalent to the gauge transformation law for the gauge fields
$\omega_{\mu a}$ in (\ref{compredtransfs}).

Note that the mixing of
spacetime and internal symmetries through the matrix $(\theta^{\mu
  a})$ determines a linear isomorphism between the frame and tangent
bundles of $\real^d$. The quantities (\ref{emunudef}) can thereby be
identified with frame fields on spacetime. This identification is
consistent with the gauge transform
\beq
\delta^\star _\ell e_\mu^\nu=X_\ell(e_\mu^\nu)-e_\mu^\lambda\,
\partial_\lambda X_\ell^\nu
\label{dbeingaugetransf}\eeq
that follows from (\ref{linmapVectRd}) and
(\ref{compredtransfs}). This is the anticipated behaviour of a
frame field under infinitesimal diffeomorphisms of $\real^d$. The
field (\ref{emunudef}) is in fact a perturbation of the usual flat
geometry of $\real^d$ with
$e_\mu^\nu\big|_{\theta=0}=\delta^\mu_\nu$. Gravitational degrees of
freedom thus arise entirely as a consequence of the noncommutative
deformation.

All of the natural geometrical objects of spacetime are thereby
canonically encoded into the noncommutative gauge fields. Let us now
consider the structure of the reduced field strength tensor
(\ref{curvcompsred}). Introduce the contractions
\beq
T_{\mu\nu}{}^\lambda~:=~-\theta^{\rho a}\,E_\rho^\lambda\,\Omega_{\mu\nu
  a}~=~E_\rho^\lambda\,\big(\nabla_\mu e_\nu^\rho-\nabla_\nu e_\mu^\rho
\big) \ ,
\label{torsiondef}\eeq
where $E_\nu^\mu$ are the inverse frame fields, i.e.
$E_\mu^\lambda\,e_\lambda^\nu=e_\mu^\lambda\,E_\lambda^\nu=\delta_\mu^\nu$,
with the formal asymptotic expansion
\beq
E_\nu^\mu=\delta_\nu^\mu+\theta^{\mu a}\,\omega_{\nu a}+
\sum_{n=2}^\infty\,\theta^{\mu a_1}\,\theta^{\mu_1a_2}\cdots
\theta^{\mu_{n-1}a_n}\,\omega_{\mu_1a_1}\cdots
\omega_{\mu_{n-1}a_{n-1}}\,\omega_{\nu a_n}
\label{inversedbeinexp}\eeq
and with the infinitesimal gauge transformation property
\beq
\delta^\star _\ell E_\nu^\mu=-X_\ell(E_\nu^\mu)-E_\lambda^\mu\,
\partial_\nu X_\ell^\lambda \ .
\label{invdbeingaugetransf}\eeq
From (\ref{dbeingaugetransf}) and (\ref{invdbeingaugetransf}) it
follows that the curvatures (\ref{torsiondef}) obey homogeneous gauge
transformation laws
\beq
\delta^\star _\ell T_{\mu\nu}{}^\lambda=X_\ell\big(T_{\mu\nu}{}^\lambda
\big) \ .
\label{deltaelltorsion}\eeq
They naturally arise as the commutation coefficients in the closure of
the commutator of covariant derivatives (\ref{nablamudef}) to a Lie
algebra with respect to the given orthonormal basis of the frame
bundle,
\beq
[\nabla_\mu,\nabla_\nu]=T_{\mu\nu}{}^\lambda~\nabla_\lambda \ .
\label{nablacommT}\eeq
The operators (\ref{nablamudef}) thereby define a non-holonomic basis
of the tangent bundle with non-holonomicity given by the
noncommutative field strength tensor. The change of basis
$\nabla_\mu=e_\mu^\nu\,\partial_\nu$ between the coordinate and
non-coordinate frames is defined by the noncommutative gauge field.

The commutation relation (\ref{nablacommT}) identifies
$T_{\mu\nu}{}^\lambda$, or equivalently the noncommutative gauge field
strengths $\Omega_{\mu\nu a}$, as the torsion tensor fields of vacuum
spacetime induced by the presence of a gravitational field. The
non-trivial frame field (\ref{emunudef}) induces a teleparallel
structure on spacetime through the linear Weitzenb\"ock connection
\beq
\Sigma_{\mu\nu}^\lambda=E_\rho^\lambda\,\nabla_\mu e_\nu^\rho \ .
\label{Weitzenbockdef}\eeq
The connection (\ref{Weitzenbockdef}) satisfies the absolute
parallelism condition $D_\mu^\Sigma e_\nu^\lambda=0$, where
$D_\mu^\Sigma$ is the Weitzenb\"ock covariant derivative. This means
that the frame fields define a mutually parallel system of local
vector fields in the tangent spaces of $\real^d$ with respect to the
tangent bundle geometry induced by (\ref{Weitzenbockdef}). The
Weitzenb\"ock connection has non-trivial torsion given by
(\ref{torsiondef}),
$T_{\mu\nu}{}^\lambda=\Sigma^\lambda_{\mu\nu}-\Sigma^\lambda_{\nu\mu}$,
but vanishing curvature, $R^\lambda{}_{\rho\mu\nu}(\Sigma)=0$. The
teleparallel structure is related to a Riemannian structure on
spacetime through the identity
\beq
\Sigma^\lambda_{\mu\nu}=\Gamma_{\mu\nu}^\lambda+K_{\mu\nu}^\lambda \ ,
\label{SigmaGammaKid}\eeq
where $\Gamma_{\mu\nu}^\lambda=\Gamma_{\mu\nu}^\lambda(e,E)$ is the
torsion-free Levi-Civita connection of the tangent bundle and
$K_{\mu\nu}^\lambda$ is the contorsion tensor. The torsion
$T_{\mu\nu}{}^\lambda$ measures the noncommutativity of displacements
of points in the flat spacetime $\real^d$, and it is dual to the
Riemann curvature tensor $R^\lambda{}_{\rho\mu\nu}(\Gamma)$ which
measures the noncommutativity of vector displacements in a curved
spacetime. Teleparallelism in this way attributes gravitation to
torsion, rather than to curvature as in general relativity, and the
Weitzenb\"ock geometry is complementary to the usual riemannian
geometry.

We have thus found that the gauge fields of the dimensionally reduced
noncommutative Yang-Mills theory naturally map onto a Weitzenb\"ock
structure, yielding an effective noncommutative field theory of
gravitation induced on flat spacetime. After some
calculation~\cite{LangSz1} one finds that the low-energy dynamics of
the dimensionally reduced noncommutative gauge theory (\ref{SNCYMdef})
is described by the local lagrangian
\beq
L_0=\frac{\big|{\rm Pf}(\theta)\big|^{-1/d}}
{4g^2}\,\det(e)\,\eta^{\mu\rho}\,\Big(\eta^{\nu\delta}\,
\eta_{\lambda\sigma}\,T_{\mu\nu}{}^\lambda\,T_{\rho\delta}{}^\sigma-
T_{\mu\nu}{}^\nu\,T_{\rho\delta}{}^\delta+2\,T_{\mu\nu}{}^\delta\,
T_{\rho\delta}{}^\nu\Big) \ ,
\label{lowendimredNCGT}\eeq
where $\eta^{\mu\nu}$ is a constant metric on $\real^d$. The Planck scale
$\kappa$ of the $d$-dimensional spacetime is given in terms the
Yang-Mills coupling constant $g$ and the noncommutativity scale as
\beq
\kappa=g\,\big|{\rm Pf}(\theta)\big|^{1/2d} \ .
\label{indPlanckscale}\eeq
The induced gravitational constant (\ref{indPlanckscale}) vanishes in
the commutative limit $\theta\to0$ and agrees in four dimensions with
that of the supergravity dual of noncommutative Yang-Mills
theory~\cite{SUGRAdual4D}. By using the relation
(\ref{SigmaGammaKid}), the lagrangian (\ref{lowendimredNCGT})
can be expressed entirely in terms of the Levi-Civita connection
$\Gamma^\lambda_{\mu\nu}$ alone, and up to a total derivative it
coincides with the standard Einstein-Hilbert lagrangian
\beq
L_{\rm E}=-\frac1{\kappa^2}\,\det(e)\,R(\Gamma)
\label{EHlagr}\eeq
in the first-order Palatini formalism. The induced lagrangian
(\ref{lowendimredNCGT}) thus defines the teleparallel formulation of
general relativity which is completely equivalent to Einstein gravity
(in the absence of spinning matter fields).

In general, there are higher-derivative corrections to the local
lagrangian (\ref{lowendimredNCGT})~\cite{LangSz1} (equivalently to
(\ref{EHlagr})). They will be treated more systematically in the next
section. These terms can be attributed to stringy
corrections to the teleparallel theory of gravity, such as those which
would arise from the trivial dimensional reduction taking the gauge
theory on a D$(2d)$-brane wrapping $\real^{2d}$ to a field theory on a
lower-dimensional D$d$-brane realized as a noncommutative soliton in
the worldvolume of the D$(2d)$-brane (see~\cite{HarveyKomaba,Szrev4}
for reviews of noncommutative solitons in this context). In these
latter instances the map (\ref{linmapVectRd}) is not surjective and
its image consists of only Poisson vector fields, satisfying ${\rm
  div}(X_\ell)=\partial_\mu X_\ell^\mu=0$, as in our
earlier analysis of this section. These higher-curvature terms
conspire, along with those induced by integrating out the auxiliary
``internal'' gauge fields $\xi_a(x)$ in (\ref{AYMconsred}), to induce
the requisite local Lorentz invariance absent in the lagrangian
(\ref{lowendimredNCGT}). For more details of these and other
teleparallel formulations in this context, see~\cite{LangSz1}.

The D-brane picture can also be used to
understand the breakdown of general covariance in the usual
noncommutative gauge theories. An infinitesimal coordinate
transformation $\delta^\star_\ell x^\mu=X_\ell^\mu(x)$ implies that the
noncommutativity parameters $\theta^{\mu a}=[x^\mu,y^a]_\star $ must
transform under gauge transformations as
\beq
\delta^\star_\ell\theta^{\mu a}~=~\big[X_\ell^\mu\,,\,y^a\big]_\star ~=~
\theta^{\nu a}\,\partial_\nu X_\ell^\mu \ .
\label{thetagaugetransf}\eeq
Requiring that the noncommutative gauge symmetries preserve the
supergravity background on the D-branes sets (\ref{thetagaugetransf})
to zero, again implying that $X$ must be a Poisson vector field.
These results all suggest that general covariance can only be achieved
in the generic settings when one considers all possible types of
noncommutativity parameter functions
$\theta=\theta(x)$~\cite{Doplicher1}. The resulting noncommutative
spaces are related to the dynamics of D-branes in curved spacetimes
and in non-constant $B$-fields. The matrix model of
Section~\ref{Matrices} naturally sums over all such D-brane
backgrounds in the form of classical vacua~\cite{Supermatrix}
representing Ricci-flat riemannian manifolds~\cite{HKK1}. Such curved
backgrounds are the topic of Section~\ref{NCGTCurved}.

\newsection{Twisted Symmetries\label{TwistSym}}

In this section we will develop an alternative approach to
implementing diffeomorphism invariance in generic noncommutative field
theories which is more universal in that it does not rely on any of
the reductions of the previous section. The idea is to {\it deform} or
``twist'' the desired spacetime symmetry in such a way that it acts
consistently on the noncommutative space, leaving the space invariant
while reducing to the standard symmetry on slowly-varying fields. In
general, the star-product of two fields will fail to be invariant
under a given symmetry transformation, as we saw in the previous
section. We will therefore keep the transformations $\delta_Xf$ of
fields $f$ intact, but deform the way in which they act on
star-products of fields. We obtain in this way twisted Leibniz rules
which can be interpreted as resulting from application of the ordinary
Leibniz rule but taking into account the transformation of the
star-product. This leads to new quantum group symmetries of
noncommutative field theories. We will first describe the general
formalism in an abstract way that can be applied later on to more
complicated noncommutative spaces. Then we study these twisted
spacetime symmetries in the simple example of the Moyal space. Among
other things, this will extend the gravitational field theory of
Section~\ref{Teleparallel} to a noncommutative one based on a quantum
deformation of the diffeomorphism group ${\rm Diff}(\real^D)$. The
relevant background on quantum groups can be found in the
book~\cite{QG1}.

\subsection{Hopf Algebras and Twist Deformations\label{TwistDefGen}}

Let $H$ be a Hopf algebra over $\complex$ with associative unital algebra
multiplication $m:H\otimes H\to H$ denoted $g\otimes h\mapsto
g\,h$. Denote the identity map on $H$ by $\id:H\to H$ and the unit
element of $H$ by $1_H$. The bialgebra structure on $H$ is implemented
by a coproduct $\Delta$ given as an algebra homomorphism
\beq
\Delta\,:\,H~\longrightarrow~H\otimes H
\label{Deltagenmap}\eeq
which is coassociative,
\beq
(\Delta\otimes\id)\circ\Delta=(\id\otimes\Delta)\circ\Delta \ .
\label{Deltacoass}\eeq
The counit $\varepsilon$ is an algebra homomorphism
$\varepsilon:H\to\complex$ obeying
\beq
(\id\otimes\varepsilon)\circ\Delta~=~\id~=~
(\varepsilon\otimes\id)\circ\Delta \ .
\label{counitDeltarels}\eeq
Finally, the antipode $S$ is an algebra anti-homomorphism $S:H\to H$
satisfying
\beq
m\circ(\id\otimes S)\circ\Delta~=~1_H\circ\varepsilon~=~
m\circ(S\otimes\id)\circ\Delta \ .
\label{antipodeDeltarels}\eeq

An invertible element $\mathcal{F}\in H\otimes H$ is said to be a
{\it twist} if it satisfies~\cite{Resh1}
\bea
(\mcF\otimes1_H)\,(\Delta\otimes\id)\mcF&=&(1_H\otimes\mcF)\,(\id\otimes
\Delta)\mcF \ , \nonumber\\ (\varepsilon\otimes\id)\mcF&=&1_{H\otimes H}~=~
(\id\otimes\varepsilon)\mcF \ .
\label{twistconds}\eea
These two conditions imply that $\mcF$ is a counital two-cocycle of the
Hopf algebra $H$. A twist element $\mcF$ determines a new Hopf algebra
structure on $H$ with twisted coproduct $\Delta_\mcF$ defined by
\beq
\Delta_\mcF(h)~=~{\rm Ad}_\mcF\circ\Delta(h)~=~
\mcF\,\Delta(h)\,\mcF^{-1}
\label{twistedcoprod}\eeq
and twisted antipode $S_\mcF$ defined by
\beq
S_\mcF(h)~=~{\rm Ad}_u\circ S(h)~=~
u\,S(h)\,u^{-1} \qquad \mbox{with} \qquad
u~=~m\circ(\id\otimes S)\mcF
\label{twistedantipode}\eeq
for $h\in H$. The resulting Hopf algebra $H_\mcF$ called a {\it
  twisted Hopf algebra}. It has the same underlying algebra structure
$m_\mcF=m$ and counit $\varepsilon_\mcF=\varepsilon$ as $H$.

Suppose now that $H$ acts on an associative, unital algebra $\alg$ over
$\complex$ with product map $\mu:\alg\otimes\alg\to\alg$ and
unit~$1_\alg$. This means that there is a bilinear map
\beq
H\otimes\alg~\longrightarrow~\alg \ , \quad
h\otimes f~\longmapsto~h\triangleright f
\label{Halgaction}\eeq
which is compatible with the algebra structure of $H$,
\beq
(h\,h')\triangleright f~=~h\triangleright(h'\triangleright f) \qquad
\mbox{and} \qquad 1_H\triangleright f~=~f \ ,
\label{Hactalgcomp}\eeq
and also with the coalgebra structure on $H$,
\beq
h\triangleright\mu(f\otimes f')~=~\mu\circ\Delta(h)\triangleright(f
\otimes f') \qquad \mbox{and} \qquad h\triangleright1_\alg~=~1_\alg\circ
\varepsilon(h) \ ,
\label{Hactcoalgcomp}\eeq
for all $h,h'\in H$ and $f,f'\in\alg$. In (\ref{Hactcoalgcomp}) the
action (\ref{Halgaction}) is extended to tensor products as $(h\otimes
h')\triangleright(f\otimes f'):=(h\triangleright
f)\otimes(h'\triangleright f')$. The first property in
(\ref{Hactcoalgcomp}) can be succinctly summarized by saying that
there is a commutative diagram
\beq
\xymatrix{f\otimes f'\ar[d]_{\mu~} ~
\ar[r]^{\!\!\!\!\!\!\!\!\!\!\!\!\!\!\Delta(h)
\triangleright}&~\Delta(h)
\triangleright(f\otimes f')\ar[d]^{~\mu} \\
\mu(f\otimes f')~\ar[r]_{\!\!\!\!\!\!\!\!h\triangleright}
&~h\triangleright\mu(f\otimes f') \ . }
\label{compcondcommdiag}\eeq
This illustrates the fact that the coproduct implements the
representation of the Hopf algebra $H$ on the tensor product
$\alg\otimes\alg$. In this case $H\subset{\rm Aut}(\alg)$, as $H$
preserves the product $\mu$ and thus acts by automorphisms of
$\alg$. If no such coproduct $\Delta$ exists, then $H$ does not act on
$\alg$.

If in addition $H$ admits a twist element $\mcF\in H\otimes H$, then
the twisted Hopf algebra $H_\mcF$ acts on the {\it twist deformed algebra}
$\alg_\mcF:=(\alg\,,\,\mu_\mcF)$ with twisted algebra product
$\mu_\mcF:\alg_\mcF\otimes\alg_\mcF\to\alg_\mcF$ defined by
\beq
\mu_\mcF(f\otimes f')=\mu\circ\mcF^{-1}\triangleright(f\otimes f')
\label{twistedalgprod}\eeq
for $f,f'\in\alg$. The first (cocycle) condition of (\ref{twistconds})
implies that the twisted product $\mu_\mcF$ is associative, while the
second (counital) condition guarantees that $1_\alg$ is an identity
element for $\mu_\mcF$, i.e. $\mu_\mcF(1_\alg\otimes
f)=f=\mu_\mcF(f\otimes1_\alg)$ for all $f\in\alg$. Using the
definition (\ref{twistedcoprod}) along with (\ref{Hactcoalgcomp}) and
(\ref{twistedalgprod}), one readily checks the requisite covariance
condition 
\bea
h\triangleright\mu_\mcF(f\otimes f')&=&h\triangleright\mu\circ
\mcF^{-1}\triangleright(f\otimes f')\nonumber\\
&=&\mu\circ\Delta(h)\,\mcF^{-1}\triangleright(f\otimes f')
\nonumber\\ &=&\mu\circ\mcF^{-1}\,\Delta_\mcF(h)\triangleright
(f\otimes f')~=~\mu_\mcF\circ\Delta_\mcF(h)\triangleright
(f\otimes f') \ .
\label{twistcovcond}\eea

This formalism gives us a new perspective on deformations in the case
of the action of a group $G$ of symmetries of an algebra $\alg$. In
this case the group algebra $H=\complex G$ is a cocommutative Hopf
algebra with coproduct $\Delta(g)=g\otimes g$, counit
$\varepsilon(g)=1$, and antipode $S(g)=g^{-1}$ for all $g\in G$
(extended to all of $\complex G$ by linearity). The antipode thus
implements the action of dual group elements on $\alg$. Then
(\ref{Hactcoalgcomp}) is just the expected covariance condition on the
multiplication
\beq
\mu\big((g\triangleright f)\otimes(g\triangleright f')\big)=
g\triangleright\mu\big(f\otimes f'\big) \ .
\label{covcondproduct}\eeq
A twist element $\mcF\in\complex G\otimes\complex G$ generically
defines a quantum deformation $G_\mcF$ of the symmetry group $G$,
generalizing the example considered in Section~\ref{NCGaugeGp}.

In what follows we will be primarily interested in the case of
infinitesimal symmetry transformations generated by the action of a
Lie algebra $\mfg$ on $\alg$. In this case the universal enveloping
algebra $H=U(\mfg)$ is a cocommutative Hopf algebra defined for any
element $X\neq1$ of $U(\mfg)$ by
\bea
\Delta(X)&=&X\otimes1+1\otimes X \ , \quad
\Delta(1)~=~1\otimes1  \ , \nonumber\\ \varepsilon(X)&=&0 \ ,
\qquad\qquad\quad\quad\quad \ 
\varepsilon(1)~=~1 \ , \nonumber\\ S(X)&=&-X \ , \qquad\qquad\qquad
S(1)~=~1 \ .
\label{UEAHopfalg}\eea
The coproduct in (\ref{UEAHopfalg}) satisfies the bialgebra property
\beq
\big[\Delta(X)\,,\,\Delta(X')\big]=\Delta\big([X,X']\big) \ ,
\label{UEAbialgebra}\eeq
and (\ref{Hactcoalgcomp}) is just the Leibniz rule
\beq
X\triangleright\mu\big(f\otimes f'\big)=\mu\big((X\triangleright
f)\otimes f'\big)+\mu\big(f\otimes(X\triangleright f')\big) \ .
\label{Lebnizcondproduct}\eeq
Any twist element $\mcF\in U(\mfg)\otimes U(\mfg)$ preserves the
commutation relations of the Lie algebra $\mfg$ and generically
defines a noncocommutative Hopf algebra $U_\mcF(\mfg)$.

In this paper we will be specifically interested in the application of
this abstract construction to the concrete example of the algebra of
functions $\alg=\alg_0=\CC^\infty(\real^D)$ with the commutative
pointwise product $\mu=\mu_0$ and with $\mfg$ a Lie algebra of spacetime
symmetries. The above construction can then be used to build
noncommutative field theories on $\real^D$ which are covariant under these
symmetries in a way which utilizes only the quantum group properties
of $U(\mfg)$. In the remainder of this section we will illustrate this
procedure in the simplest case of field theories on Moyal space.

\subsection{Twisted Poincar\'e Transformations\label{TwistPoincare}}

Consider the Poincar\'e algebra
$\mfg=\mathfrak{iso}(D-1,1)=\mathfrak{so}(D-1,1)\ltimes\real^D$ in $D$
dimensions with translation generators $P_i$ and Lorentz
generators $M_{ij}=-M_{ji}$, $i,j=1,\dots,D$ obeying the commutation
relations
\bea
[P_i,P_j]&=&0 \ , \nonumber\\
{}[M_{ij},M_{kl}]&=&\eta_{ik}\,M_{jl}-\eta_{il}\,M_{jk}-\eta_{jk}\,
M_{il}+\eta_{jl}\,M_{ik} \ , \nonumber\\
{}[M_{ij},P_k]&=&\eta_{ik}\,P_j-\eta_{jk}\,P_i \ .
\label{Poincarealg}\eea
Acting on the commutative algebra $\alg_0(\real^D)$, we can represent
these generators by the usual linear and angular momentum operators
\beq
P_i~=~\partial_i \qquad \mbox{and} \qquad
M_{ij}~=~x_i\,\partial_j-x_j\,\partial_i \ .
\label{PiMijcommrep}\eeq
Equip the universal enveloping algebra $H=U(\mfg)$
with the primitive coproduct $\Delta_0:=\Delta$ defined in
(\ref{UEAHopfalg}). Let $\theta$ be a constant Poisson structure on
$\real^D$. Taking the exponential of the corresponding Poisson tensor
we introduce the abelian Drin'feld twist element
$\mcF_\theta=\exp\big(-\frac\ii2\,\theta^{ij}\,P_i\otimes P_j\big)$ as
in (\ref{Moyaltwistdef}).

The corresponding twisted Hopf algebra $H_\theta:=H_{\mcF_\theta}$
acts on the Moyal space described in Section~\ref{MoyalProd}. The
Moyal product $\mu_\theta:=\mu_{\mcF_\theta}=\mu_0\circ\mcF_\theta^{-1}$ is a
bidifferential operator. The abelian twist $\mcF_\theta$ leaves the
commutation relations (\ref{Poincarealg}) unchanged but deforms the
bialgebra structure of $H$. By using the Hadamard formula
\beq
{\rm Ad}_{\e^X}(Y)~=~\e^X\,Y~\e^{-X}~=~\sum_{n=0}^\infty\,\frac1{n!}\,
[\,\underbrace{X,[X,\dots[X}_n\,,Y]\dots]]~=~
\sum_{n=0}^\infty\,\frac{({\rm ad}_X)^n}{n!}\,Y
\label{Hadamard}\eeq
along with (\ref{Poincarealg}), the twisted
coproducts $\Delta_\theta:=\Delta_{\mcF_\theta}$ of the Poincar\'e
generators are straightforwardly computed with the
result~\cite{CKNT1,Wess1}
\bea
\Delta_\theta(P_i)&=&P_i\otimes1+1\otimes P_i \ , \nonumber\\
\Delta_\theta(M_{ij})&=&M_{ij}\otimes1+1\otimes M_{ij} \nonumber\\
&& +\,\mbox{$\frac\ii2$}\,
\theta^{kl}\,\big((\eta_{ik}\,P_j-\eta_{jk}\,P_i)\otimes P_l+
P_k\otimes(\eta_{il}\,P_j-\eta_{jl}\,P_i)\big) \ .
\label{Poincarecoprods}\eea

It follows that while the translation operators $P_i$ satisfy the
usual Leibniz rule with respect to the star-product
(\ref{Moyalprodtwistdef}), the Lorentz generators $M_{ij}$ in
(\ref{PiMijcommrep}) are not derivations of the Moyal product. This
simply reflects the fact that noncommutative field theory on Moyal
spaces is translationally invariant but not Lorentz invariant. In the
string theory setting, the loss of Lorentz invariance is due to the
{\it fixed} expectation value of the supergravity $B$-field. The field
theory is invariant under ``observer'' Lorentz
transformations~\cite{Kostobserver},
i.e. rotations or boosts of observer inertial frames. As discussed in
Section~\ref{StarGaugeGeom}, in this case the covariant transformation
of $\theta^{ij}$ can be gauged away by a star-gauge transformation, as
exhibited for instance in (\ref{thetagaugetransf}). However, the
field theory is not invariant under ``particle''
Lorentz transformations, leaving $\theta^{ij}$ invariant, which
correspond to rotations or boosts of localized field configurations
within a {\it fixed} observer frame of reference. 

Canonical noncommutative field theory is, however, invariant under
{\it twisted} particle transformations~\cite{CKNT1}. We can see this by writing
down the actions of the Poincar\'e generators (\ref{PiMijcommrep}) on
the noncommutative algebra $\alg_\theta:=\alg_{\mcF_\theta}$ in such a
way that the twisted Poincar\'e transformations are compatible with
the Moyal product. For this, we need to examine how the pointwise
product of two functions $f$ and $g$ is represented in
$\alg_\theta$~\cite{Wessgrav1}. Consider the identity
\beq
f\,g~=~\mu_0\circ\big(\mcF_\theta^{-1}\,\mcF_\theta^{\phantom{1}}\big)
\triangleright\big(f\otimes g\big)~=~\mu_0\circ
\mcF_\theta^{-1}\triangleright
\big(\mcF_\theta^{\phantom{1}}\triangleright(f\otimes g)\big) \ .
\label{ptwiseobvious}\eeq
Expand the second exponential $\mcF_\theta$ and use the first
exponential $\mcF_\theta^{-1}$ to write each term as a
star-product. In this way we can represent $f\,g$ as a formal
asymptotic expansion in star-products of $f$, $g$ and their
derivatives as
\beq
f\,g=f\star g+\sum_{n=1}^\infty\,\left(-\frac\ii2\right)^n\,
\frac1{n!}\,\theta^{i_1j_1}\cdots\theta^{i_nj_n}\,
(\partial_{i_1}\cdots\partial_{i_n}f)\star
(\partial_{j_1}\cdots\partial_{j_n}g) \ .
\label{fgfstarg}\eeq

From (\ref{fgfstarg}) it follows that the actions of the vector fields
(\ref{PiMijcommrep}) on $\alg_\theta$ are given by
\bea
P_i^\star\triangleright f&:=&\partial_i^\star\triangleright f~=~
\partial_if \ , \nonumber\\ M_{ij}^\star\triangleright f&=&
\big(x_i\,\partial_j-x_j\,\partial_i-\mbox{$\frac\ii2$}\,
(\theta_i{}^k\,\partial_j\,\partial_k-\theta_j{}^k\,\partial_i\,
\partial_k)\big)f \ ,
\label{PiMijstardef}\eeq
where we have used $f\star1=f=1\star f$. Thus the first order
differential operator $M_{ij}$ on $\alg_0$ becomes a second order
differential operator $M_{ij}^\star$ on $\alg_\theta$, reflecting the
presence of the extra terms in the corresponding twisted coproduct in
(\ref{Poincarecoprods}). The additional terms are required to make the
twisted Lorentz transformations compatible with the star-product on
$\alg_\theta$. In particular, these symmetries transform coordinates
into derivatives (momenta) according to
\beq
M^\star_{ij}\triangleright x^k=\delta_j{}^k\,x_i-\delta_i{}^k\,x_j-
\mbox{$\frac\ii2$}\,\big(\theta_i{}^l\,\delta_j{}^k-\theta_j{}^l\,
\delta_i{}^k\big)\,\partial_l \ ,
\label{coordtoderiv}\eeq
which coincides with the usual Lorentz transformation for
$\theta=0$. This illustrates the inherent non-locality of the twisted
spacetime symmetry transformations.

One easily checks that (\ref{PiMijstardef}) gives a representation of
the Poincar\'e algebra (\ref{Poincarealg}) on
$\alg_\theta$. Furthermore, using (\ref{fgfstarg}) one confirms the
expected covariance of fields in $\alg_\theta$ under twisted
Poincar\'e transformations~\cite{CKNT1,Wessgrav1}. Finally, from either
(\ref{Poincarecoprods}) or (\ref{PiMijstardef}) one straightforwardly
computes the Lorentz transform of the Moyal bracket of coordinate
generators with the result
\beq
M_{kl}^\star\triangleright\big[x^i\,,\,x^j\big]_\star~=~
\mu_\theta\circ\Delta_\theta(M_{kl})\triangleright\big(
x^i\otimes x^j-x^j\otimes x^i\big)~=~0 \ .
\label{bracketLorentz}\eeq
From (\ref{Heisenalg}) and (\ref{bracketLorentz}) it follows that
$M_{kl}^\star\triangleright\theta^{ij}=0$, and so the antisymmetric
tensor $\theta^{ij}$ is invariant under twisted Lorentz
transformations, i.e. twisted spacetime symmetries induce particle
transformations~\cite{G-BLR-RV1}. Due to the comultiplication rule
(\ref{Poincarecoprods}) one doesn't need to transform $\theta$, as was
done in Section~\ref{StarGaugeGeom}, to obtain Poincar\'e covariance.

This twisted Poincar\'e covariance has important ramifications for
relativistic noncommutative quantum field theory~\cite{CKNT1,CPT1},
because it extends the naive symmetry group which preserves $\theta$
to the full Poincar\'e symmetry group. Since the commutation relations
(\ref{Poincarealg}) are unaffected by the twist, noncommutative fields
can still be characterized according to their transformation
properties under the Lorentz group. In addition, representations of
the twisted Poincar\'e algebra are classified, just as in the
commutative case, by ordinary mass and spin eigenvalues. Thus the
entire representation theoretic content of noncommutative quantum
field theory is identical to that of the corresponding commutative
theory with the usual Poincar\'e symmetry. This leads to
noncommutative versions of many of the standard theorems of
relativistic quantum field theory such as the CPT theorem, the
spin-statistics theorem, and Haag's theorem, among
others~\cite{CPT1}. Further physical consequences of the twisted
Poincar\'e symmetry of noncommutative quantum field theory are
explored in~\cite{Bal1,Bal2,Gang1,Zahn1}. The global version, i.e. the
twisted action of the Poincar\'e group, is described in~\cite{GKMG1}.

\subsection{Twisted Diffeomorphisms\label{TwistDiff}}

The analysis of Poincar\'e symmetries above generalizes in a
straightforward and systematic way to the Lie algebra $\Vect(\real^D)$
of infinitesimal diffeomorphisms. Much of what we have said above
concerning covariance carries through to give the notion of twisted
general covariance. The twisted diffeomorphisms again act as particle
transformations, leaving the Poisson structure $\theta$ on $\real^D$
invariant. However, now invariance under observer transformations is
lost in general, as the generic transformation of $\theta^{ij}$ will
lead to a different class of noncommutative spacetimes as discussed in
Section~\ref{Teleparallel} (see (\ref{thetagaugetransf})). These
spaces are the topic of the next section. We will now describe these
generic twisted spacetime symmetries and use them to systematically
compute the higher derivative corrections to the Einstein-Hilbert
lagrangian (\ref{EHlagr}) arising from canonical noncommutativity. As
before, the twist does not change the action of infinitesimal
diffeomorphisms on fields, only the coproduct and consequently the
action of diffeomorphisms on star-products. Extensions of the twisted
Poincar\'e symmetry are described
in~\cite{Oeckl1,Matlock1,LSV1,Bal3,Wessgrav1,Wessgrav2}.

The Lie algebra $\mfg=\Vect(\real^D)$ is generated by vector fields
\beq
X=X^i(x)~\mbox{$\frac\partial{\partial x^i}$}
\label{VectRDgens}\eeq
with polynomial coefficient functions $X^i$ on $\real^D$. The Lie
bracket of two vector fields $X$ and $Y$ is given by
\beq
[X,Y]~=~[X,Y]^i(x)~\mbox{$\frac\partial{\partial x^i}$} \qquad
\mbox{with} \qquad [X,Y]^i~=~X^j\,\partial_jY^i-Y^j\,\partial_jX^i \ .
\label{VectRDbracket}\eeq
Work in the enveloping algebra $U(\mfg)$. Then the twisting procedure
of Section~\ref{TwistDefGen} above gives a prescription for encoding
the action of arbitrary differential operators, of any order, with
polynomial coefficients on Moyal
products~\cite{G-BLR-RV1,Wessgrav1}. By using the Hadamard
formula (\ref{Hadamard}), one computes the twisted coproduct of an
arbitrary vector field (\ref{VectRDgens}) as the formal asymptotic
expansion
\bea
\Delta_\theta(X)&=&X\otimes1+1\otimes X \nonumber\\ &&
+\,\sum_{n=1}^\infty\,\left(-\frac\ii2\right)^n\,\frac1{n!}\,
\theta^{i_1j_1}\cdots\theta^{i_nj_n}\,\bigl([\partial_{i_1},[
\partial_{i_2},\dots[\partial_{i_n},X]\dots]]\otimes\partial_{j_1}
\partial_{j_2}\cdots\partial_{j_n} \bigr. \nonumber \\ &&
\qquad\qquad\qquad\qquad +\bigl.\, \partial_{i_1}\partial_{i_2}
\cdots\partial_{i_n}\otimes[\partial_{j_1},[\partial_{j_2},\dots[
\partial_{j_n},X]\dots]]\bigr) \ .
\label{coproddifftwist}\eea
This twisted coproduct defines the action of the Lie algebra
(\ref{VectRDbracket}) of vector fields on the star-product of two
fields. Unlike the standard Leibniz rule, it guarantees that the Moyal
product transforms covariantly under twisted diffeomorphisms.

From (\ref{fgfstarg}) it follows that the action of a vector field
(\ref{VectRDgens}) on the noncommutative algebra
$\alg_\theta(\real^D)$ is given by the asymptotic series
\beq
X^\star\triangleright f=X(f)+\sum_{n=1}^\infty\,\left(-\frac\ii2
\right)^n\,\frac1{n!}\,\theta^{i_1j_1}\cdots\theta^{i_nj_n}\,
\big(\partial_{i_1}\cdots\partial_{i_n}X^i\big)\star\big(
\partial_{j_1}\cdots\partial_{j_n}\partial_if\big) \ .
\label{Xstarfdef}\eeq
Thus a vector field on $\real^D$ becomes a higher-order
differential operator acting on fields $f\in\alg_\theta$. One verifies
that the operators $X^\star$ represent the Lie algebra
(\ref{VectRDbracket}) of vector fields as
\beq
\big[X^\star\,,\,Y^\star\big]_\star\triangleright f=
\big[X\,,\,Y\big]^\star\triangleright f \ .
\label{Xstarbracket}\eeq

A generic tensor field $T^{i_1\cdots i_p}_{j_1\cdots j_q}$ on
$\real^D$ of rank $(p,q)$ transforms under twisted diffeomorphisms
as~\cite{Wessgrav1}
\bea
\delta_X^\star T^{i_1\cdots i_p}_{j_1\cdots j_q}&=&
-X^\star\triangleright T^{i_1\cdots i_p}_{j_1\cdots j_q}-
\big(\partial_{j_1}X^k\big)^\star\triangleright T^{i_1\cdots i_p}_{kj_2
\cdots j_q}-\dots-\big(\partial_{j_q}X^k\big)^\star\triangleright
T^{i_1\cdots i_p}_{j_1\cdots j_{q-1}k} \nonumber\\ && +\,
\big(\partial_kX^{i_1}\big)^\star\triangleright
T^{ki_2\cdots i_p}_{j_1\cdots j_q}+\dots+\big(\partial_kX^{i_p}
\big)^\star\triangleright T^{i_1\cdots i_{p-1}k}_{j_1\cdots j_q} \ .
\label{deltaXtensor}\eea
The twisted coproduct (\ref{coproddifftwist}) ensures that the
star-product $T^{i_1\cdots i_p}_{j_1\cdots j_q}\star T^{k_1\cdots
  k_r}_{l_1\cdots l_s}$ of two tensor fields of ranks $(p,q)$ and
$(r,s)$ transforms as a tensor field of rank $(p+r,q+s)$. For example,
given any two scalar fields $f,g\in\alg_\theta$, from the definition
(\ref{Xstarfdef}) we obtain~\cite{Wessgrav1}
\beq
\delta_X^\star(f\star g)~=~-X^\star\triangleright(f\star g)~=~-
X(f\star g) \ .
\label{deltaXfstarg}\eeq
Thus the Moyal product transforms covariantly with respect to twisted
diffeomorphisms. In this way, tensor calculus on the noncommutative
space $\real_\theta^D$ is established through representations of the
twisted Hopf algebra
$U_\theta(\Vect(\real^D)):=U_{\mcF_\theta}(\mfg)$. This fact will now be
exploited to regard $U_\theta(\Vect(\real^D))$ as the underlying
symmetry algebra of a noncommutative theory of
gravity~\cite{Wessgrav1}. The beauty behind this construction is that
it yields a deformation of general relativity which is based on a
general, underlying dynamical symmetry principle.

Let $e_i^a$, $i,a=1,\dots,D$ be classical frame fields for a metric
tensor $g_{ij}:=e_i^a\,\eta_{ab}\,e_j^b$ on $\real^D$. Define a
noncommutative metric tensor $G_{ij}$ on $\real_\theta^D$ by
\beq
G_{ij}=\mbox{$\frac12$}\,\eta_{ab}\,\bigl(e_i^a\star e_j^b+e_j^a
\star e_i^b\bigr)
\label{metricthetadef}\eeq
with the property $G_{ij}\big|_{\theta=0}=g_{ij}$. It transforms as a
symmetric tensor of rank two in $\alg_\theta$,
\beq
\delta_X^\star G_{ij}=-X^\star\triangleright G_{ij}-\big(
\partial_iX^k\big)^\star\triangleright G_{kj}-\big(\partial_jX^k
\big)^\star\triangleright G_{ik} \ .
\label{deltaXGij}\eeq
Let $G^{\star\,ij}$ denote the star-inverse of $G_{ij}$ with
\beq
G_{ij}\star G^{\star\,jk}=\delta_i{}^k \ .
\label{Gijstarinv}\eeq
The twisted Christoffel symbols $\Gamma_{ij}^k=\Gamma_{ji}^k$ can be
computed from the noncommutative metric $G_{ij}$ and its star-inverse
as
\beq
\Gamma_{ij}^k=\mbox{$\frac12$}\,\left(\partial_i^\star\triangleright
G_{jl}+\partial_j^\star\triangleright
G_{il}-\partial_l^\star\triangleright G_{ij}\right)\star
G^{\star\,lk} \ .
\label{NCChristoffel}\eeq
The corresponding twisted Ricci tensor
\beq
R_{ij}:=R_{ikj}{}^k
\label{NCRicci}\eeq
is given in terms of the twisted Riemann curvature tensor
\beq
R_{ijk}{}^l=\partial_j^\star\triangleright\Gamma_{ik}^l-
\partial^\star_i\triangleright\Gamma^l_{jk}+\Gamma_{jk}^p\star
\Gamma^l_{ip}-\Gamma_{ik}^p\star\Gamma_{jp}^l \ .
\label{NCRiemann}\eeq

A noncommutative deformation of the Einstein-Hilbert lagrangian
(\ref{EHlagr}) can now be written down in the form
\beq
L_{\rm E}^\theta=-\frac1{2\kappa^2}\,\det{}_\star(e)\star
G^{\star\,ij}\star\bigl(R_{ij}+R_{ij}{}^\dag\bigr)
\label{EHactiontheta}\eeq
with $L_{\rm E}^{\theta=0}=L^{~}_{\rm E}$, where we have defined the
star-determinant by
\beq
\det{}_\star(e)=\mbox{$\frac1{D!}$}\,\epsilon^{i_1\cdots i_D}\,
\epsilon_{a_1\cdots a_D}~e_{i_1}^{a_1}\star\cdots\star e_{i_D}^{a_D} \
.
\label{stardetdef}\eeq
Using the twisted coproduct (\ref{coproddifftwist}) one readily
computes~\cite{Wessgrav1}
\beq
\delta_X^\star L_{\rm E}^\theta=-\partial_i^\star\triangleright\bigl(
(X^i)^\star\triangleright L_{\rm E}^\theta\bigr) \ ,
\label{LEtheta}\eeq
and as a consequence of (\ref{PiMijstardef}) and (\ref{LEtheta}) the
corresponding action $S_{\rm E}^\theta:=\int\dd^Dx~L_{\rm E}^\theta$
is invariant under arbitrary twisted diffeomorphisms, $\delta_X^\star
S_{\rm E}^\theta=0$. Via an explicit asymptotic expansion in $\theta$,
one can compute the higher-derivative noncommutative corrections to
Einstein gravity described by the lagrangian
(\ref{EHlagr})~\cite{Wessgrav1}. The generically complex nature of the
twisted Ricci tensor $R_{ij}$ here is reminescent of what occurs in
other noncommutative deformations of gravity which require
complexification of the metric and of the local Lorentz symmetry
group~\cite{ComplexGrav1,ComplexGrav2,ComplexGrav3}.

However, unlike the situation with noncommutative Yang-Mills theory, it is
not clear yet how this theory of noncommutative gravity originates as
an ultraviolet completion in string theory. This difficulty may be
due to the fact that the diffeomorphism invariance that we have
achieved here is not realized as a sort of star-gauge symmetry, as we
attempted to do in the previous section. Since twisted diffeomorphisms
do not give a standard Leibniz rule, effectively producing a variation
of the star-product~\cite{Alvarezgrav1}, it is not clear whether or
not they can be implemented at the quantum level. The quantization of
systems with a twisted symmetry appears to be quite different than the
conventional quantizations. The gravitational interactions
induced on a D-brane in the presence of a constant background
$B$-field in the Seiberg-Witten decoupling limit cannot be expressed
solely in terms of Moyal products~\cite{Alvarezgrav1}, and thus string
theory contains far richer dynamics than that of the gravity lagrangian
(\ref{EHactiontheta}). The problem of writing the correct form of the
effective action for noncommutative gravity on D-branes is also
addressed in~\cite{Ghodsi1}. Until their role is clarified to the extent
that star-gauge symmetry plays for gauge theories, acting only on
fields as in the commutative theories, the role played by twisted
symmetries in the context of string theory remains clouded in
mystery.

\newsection{Noncommutative Gauge Theories on Curved
  Backgrounds\label{NCGTCurved}}

Our considerations of the previous sections naturally drive us away
from the simple Moyal spaces to more complicated noncommutative
geometries. In this section we will describe a fairly general
framework which is applicable to the dynamics of D-branes in {\it
  curved} string backgrounds. In the next section we consider a
variety of explicit examples. These generalizations require us to
bring in a host of formal techniques from the theory of deformation
quantization of generic Poisson manifolds. We will focus on those
aspects which are new to this framework as well as the role of twisted
spacetime symmetries in these more general settings.

\subsection{Kontsevich Formula\label{KontFormula}}

Besides the technical reasons described previously for wanting to extend
the framework of noncommutative gauge symmetries to more general
situations, there is a precise physical instance which can aid us in
developing the general formalism that we need. The generalized
noncommutative spaces we are interested in arise through deformations
of D-brane worldvolumes $M$, embedded in some target spacetime, in the
presence of a background $B$-field $B=\frac12\,B_{ij}(x)~\dd
x^i\wedge\dd x^j$. The D-brane will also typically carry a two-form
${\rm U}(1)$ gauge field strength $F=\dd A$, and one should consider
instead the gauge-invariant combination $\mcB:=B+F$ (We work in string
units $2\pi\,\alpha'=1$ throughout). The $B$-field has NS--NS
three-form field strength $H=\frac13\,H_{ijk}(x)~\dd x^i\wedge\dd
x^j\wedge\dd x^k$ given by
\beq
H~=~\dd B~=~\dd\mcB \ .
\label{HddB}\eeq
The curvature of the $B$-field is tied to the curvature of the metric
$g=\frac12\,g_{ij}(x)~\dd x^i\otimes\dd x^j$ of the given closed
string background. To leading order in the string length, the
beta-function equations which describe a consistent closed string
background read
\beq
R_{ij}~=~\mbox{$\frac14$}\,H_{ikl}\,H_j{}^{kl} \qquad \mbox{and}
\qquad \nabla^iH_{ijk}~=~0 \ .
\label{betaleadorder}\eeq
The effective open string metric $G$ and noncommutativity bivector field
$\theta=\frac12\,\theta^{ij}(x)\,\partial_i\wedge\partial_j$ seen by
the D-brane is given by the Seiberg-Witten matrix inversion
formula~\cite{SW1}
\beq
G+\theta=(g+\mcB)^{-1} \ .
\label{GthetagB}\eeq
The structure of the $B$-field thus controls the noncommutativity of
the effective open string dynamics.

In these generic situations the appropriate modification of the Moyal
star-product is provided by the Kontsevich
formula~\cite{CornSch1,Cycl1,Kont1} which can be expressed as the
asymptotic expansion
\bea
f\star g&=&f\,g+\mbox{$\frac\ii2$}\,\theta^{ij}\,\partial_if\,
\partial_jg-\mbox{$\frac18$}\,\theta^{ik}\,\theta^{jl}\,
\partial_i\partial_jf\,\partial_k\partial_lg \nonumber\\ &&
-\,\mbox{$\frac1{12}$}\,\theta^{il}\,\partial_l\theta^{jk}\,\bigl(
\partial_i\partial_jf\,\partial_kg-\partial_jf\,\partial_i\partial_kg
\bigr)+O\big(\theta^3\big) \ ,
\label{Kontexp}\eeq
while open string parameters are still given by the same formulas as
above. Given three generic functions $f$, $g$ and $h$, one easily
computes
\beq
(f\star g)\star h-f\star(g\star h)=\mbox{$\frac16$}\,\bigl(\theta^{il}\,
\partial_l\theta^{jk}+\theta^{jl}\,\partial_l\theta^{ki}+
\theta^{kl}\,\partial_l\theta^{ij}\bigr)\,\partial_if\,\partial_jg\,
\partial_kh+O\big(\theta^3\big) \ .
\label{fghassoc}\eeq
It follows that if the components of the bivector field $\theta$ satisfy
\beq
\theta^{il}\,\partial_l\theta^{jk}+\theta^{jl}\,\partial_l\theta^{ki}+
\theta^{kl}\,\partial_l\theta^{ij}=0 \ ,
\label{thetaPoissoncond}\eeq
then the corresponding star-product is associative (to all orders in
$\theta$). The condition (\ref{thetaPoissoncond}) can be expressed in
a global, coordinate-free form by introducing the graded
Schouten-Nijenhuis Lie bracket for polyvector fields $X=X^{i_1\cdots
  i_{k_X}}\,\partial_{i_1}\wedge\cdots\wedge\partial_{i_{k_X}}$ and
$Y=Y^{i_1\cdots
  i_{k_Y}}\,\partial_{i_1}\wedge\cdots\wedge\partial_{i_{k_Y}}$ through
\beq
[X,Y]_{\rm S}=(-1)^{k_X-1}\,X\bullet Y-(-1)^{k_X\,(k_Y-1)}\,
Y\bullet X
\label{SNbracketdef}\eeq
where
\beq
X\bullet Y:=\sum_{l=1}^{k_X}\,(-1)^{l-1}\,X^{i_1\cdots i_{k_X}}\,
\partial_lY^{j_1\cdots j_{k_Y}}\,\partial_{i_1}\wedge\cdots\wedge
\widehat{\partial_{i_l}}\wedge\cdots\wedge\partial_{i_{k_X}}\wedge
\partial_{j_1}\wedge\cdots\wedge\partial_{j_{k_Y}}
\label{XbulletYdef}\eeq
and the hat indicates an omitted derivative. Then
(\ref{thetaPoissoncond}) is equivalent to the vanishing
Schouten-Nijenhuis bracket
\beq
[\theta,\theta]_{\rm S}=0 \ .
\label{thetaSchouten}\eeq

This condition means that the bivector field $\theta$ defines a Poisson
structure on the worldvolume $M$ with Poisson bracket
\beq
\{f,g\}_\theta~=~\theta(\dd f,\dd g)~=~\theta^{ij}\,\partial_if\,
\partial_jg
\label{thetaPoissondef}\eeq
for functions $f,g\in\CC^\infty(M)$, which is the $O(\theta)$ term in
(\ref{Kontexp}). The condition (\ref{thetaSchouten}) is equivalent to
the Jacobi identity for (\ref{thetaPoissondef}). Note that the
symmetric part $G$ in (\ref{GthetagB}), i.e. the open string metric on
the brane, does not contribute to the deformation quantization,
because it defines a Hochschild cocycle. Any Hochschild coboundary can
be removed by a gauge transformation which corresponds to an
invertible differential operator
$\mcD:\CC^\infty(M)\to\CC^\infty(M)$. This is the content of the
formality theorem of Section~\ref{Formality} below, and it defines a
cohomologically {\it equivalent} star-product given by
\beq
f\star'g:=\mcD^{-1}(\mcD f\star\mcD g) \ .
\label{newstarequiv}\eeq
If $\mcD=1+D^{ij}\,\partial_i\partial_j+\dots$, then all terms of the
form $G^{ij}\,\partial_if\,\partial_jg$ can be gauged away with
$D^{ij}=-G^{ij}$.

The generalized Maxwell equations on the brane, which come from
variation of the Born-Infeld density $\sqrt{\det(g+\mcB)}$ with
respect to the gauge connection $A$, can be recast using
(\ref{GthetagB}) into the form
\beq
\partial_i\bigl(\,\sqrt{\det(g+\mcB)}~\theta^{ij}\,\bigr)=0 \ .
\label{BIeom}\eeq
Using this equation along with various worldsheet operator product
expansions and factorizations, one can show~\cite{CornSch1} that even when the
condition (\ref{thetaSchouten}) is violated, the bivector field $\theta$
still endows $\CC^\infty(M)$ with the structure of an {\it
  ${\rm A}_\infty$-algebra}. This case corresponds to the embedding of a
curved brane in a curved background, and these are the natural
algebras that appear in generic open-closed string field
theories~\cite{OpenclosedSFT}. It means that the algebra
$\CC^\infty(M)$ is endowed with an ${\rm A}_\infty$ homotopy
associative structure, whereby the failure of associativity of the
star-product is controlled by a third-order term, and similarly for
all higher orders. Thus although $(f_1\star f_2)\star f_3\neq
f_1\star(f_2\star f_3)$ in general, there is a homotopy ${\mathcal
  O}_3(f_1,f_2,f_3)(\mu_\theta):[0,1]\times\CC^\infty(M)^{\otimes
  3}\to\CC^\infty(M)$ between the two seemingly distinct ways of
grouping three functions under the star-product. This extends to
homotopies ${\mathcal
  O}_n(f_1,\dots,f_n)(\mu_\theta):[0,1]\times\CC^\infty(M)^{\otimes
  n}\to\CC^\infty(M)$ of grouping $n$ functions for all $n>3$. In
particular, it implies that the star-commutator algebra
\beq
\big[x^i\,,\,x^j\big]_\star=\theta^{ij}(x) \ ,
\label{starcommalg}\eeq
although not a Lie algebra as in the associative case, is an {\it
  ${\rm L}_\infty$-algebra}, i.e. the Jacobi identity is satisfied up
to homotopy and similarly for all higher order star-commutators.

In the following we will always assume for simplicity that $\theta$ is
a Poisson bivector field, obeying (\ref{thetaSchouten}). This corresponds to
embeddings of a curved D-brane in a flat background spacetime. The
topological limit corresponds to the situation in which the closed
string metric $g$ is much smaller than all skew-eigenvalues of
$\mcB$, so that $\theta=\mcB^{-1}$ is nondegenerate (again this
necessitates $D:=\dim(M)$ even). Then the Poisson condition
(\ref{thetaSchouten}) is equivalent to a vanishing NS--NS field
strength $H=\dd B=\dd\mcB=0$ on the D-brane and the Born-Infeld
measure reduces to the Liouville measure for the symplectic structure
$\omega:=\theta^{-1}$. The Kontsevich formula continues to work if we
drop the assumption that $\theta$ is invertible, but then the natural
measure is lost.

In the nondegenerate case, we can straightforwardly justify the use of
the Kontsevich formula (\ref{Kontexp}) in the present context directly
from the open string dynamics on the D-brane~\cite{CornSch1}, which in
the topological limit above are obtained from an A-model topological
open string theory called the {\it Poisson
  sigma-model}~\cite{CatFel1}. Let $X$ be a string embedding field
with worldsheet the upper half-plane $\Sigma=\mathbb{H}$ (or the disc)
whose boundary $\partial\Sigma=\real$ is mapped into a Poisson D-brane
worldvolume $M$. The action is defined by pullback as
\beq
S_{\rm top}~=~\int_\Sigma\,X^*\omega~=~
\int_\Sigma\,\omega_{ij}(X)~\dd X^i\wedge\dd X^j \ .
\label{Poissonsigaction}\eeq
Consider perturbation theory of the corresponding path integral around the
trivial constant classical solution $X(z,\overline{z}\,)=x\in M\quad\forall
\,(z,\overline{z}\,)\in\Sigma$. The space of trivial solutions coincides with
$M$. Let $f,g\in\CC^\infty(M)$. Let $r<s<t$ be any three cyclically
ordered points on the real line $\partial\Sigma$ defined by
$z=\overline{z}$. Then the Kontesevich formula can be expressed as the
perturbative expansion of the path integral
\beq
(f\star g)(x):=\int_{X(r)=x}\,\big[\dd\hat X
\big]~\e^{\ii\hat S_{\rm top}(\hat
  X)}~f\big(X(t)\big)\,g\big(X(s)\big)
\label{Kontcorr}\eeq
taken over the space of bundle morphisms $\hat X:T\Sigma\to T^*M$
with base map $X:\Sigma\to M$. This expression is independent of the
choice of points $r,s,t\in\real$ owing to the topological nature of
the string theory.

One can check that this expression for the star-product defines a
deformation quantization of $\CC^\infty(M)$ in the usual
sense~\cite{CatFel1}. Associativity is a consequence of the
topological nature of the
model, as the three-point correlation functions are invariant under
cyclic permutations of the fields due to the invariance of the action
under orientation-preserving diffeomorphisms of $\Sigma$. This can be
proven rigorously by using Stokes' theorem on
compactified configuration spaces. The unit element is
$f=1$. Rescaling $\theta$ by a parameter $\hbar$, it has a
formal asymptotic expansion
\beq
f\star g=f\,g+\sum_{n=1}^\infty\,(\ii\hbar)^n~{\rm B}_n(f,g)
\label{Kontformalexp}\eeq
where ${\rm B}_n$ for each $n\in\nat$ is a bidifferential operator of
degree $n$ with
\beq
{\rm B}_1(f,g)=\mbox{$\frac12$}\,\{f,g\}_\theta
\label{B1fg}\eeq
as in (\ref{Kontexp}). These expansion coefficients can be computed
explicitly through the Feynman diagram expansion of the correlation
function (\ref{Kontcorr}), reproducing the standard diagrammatic
expression for the original Kontsevich deformation quantization as a
sum over admissible graphs of order $n$ in
(\ref{Kontformalexp})~\cite{Kont1}. Other associative open string
noncommutative deformations, valid also in the non-topological limit
$H\neq0$, can be found in~\cite{HoYeh1,HoMiao1,Ho1}.

To write down a gauge theory action later on, we will need an
appropriate definition of integration. Generically, this requires the
introduction of a measure $\Omega$ on $M$ as independent input. Let
$\theta$ be a Poisson structure which is divergence-free with respect
to a measure $\Omega$ on $M$,
\beq
\partial_i\big(\Omega\,\theta^{ij}\big)=0 \ .
\label{divtheta0}\eeq
Then there exists a (cyclic) star-product which is gauge-equivalent to
the Kontsevich product and which obeys
\beq
\int_M\,\Omega~(f\star g)\,h=\int_M\,\Omega~(g\star h)\,f
\label{Kontcycl1}\eeq
for functions $f,g,h\in\CC^\infty(M)$ of compact
support~\cite{FelSh1,CalWohl1}. Setting $h=1$ and using the identity
$g\star1=g=1\star g$, this immediately implies the generalized
Connes-Flato-Sternheimer conjecture
\beq
\int_M\,\Omega~f\star g=\int_M\,\Omega~f\,g \ .
\label{Kontcycl2}\eeq

In the Poisson sigma-model, the cyclicity formula (\ref{Kontcycl1})
follows by calculating the correlation function~\cite{FelSh1}
\beq
\langle f\,g\,h\rangle:=\int\,\big[\dd\hat X
\big]~\e^{\ii\hat S_{\rm top}(\hat X)}~
f\bigl(X(t)\bigr)\,g\bigl(X(s)\bigr)\,h\bigl(X(r)\bigr) \ .
\label{fghcorrfn}\eeq
The path integral (\ref{fghcorrfn}) can be evaluated by fixing any
convex linear combination of $X(t)$, $X(s)$, $X(r)$ equal to $x\in M$
as boundary condition, and then integrating over $x$ in the
measure~$\Omega$. The result is formally independent of the choice of
linear combination. With the choices $X(r)=x$ and $X(t)=x$ we obtain
(\ref{Kontcycl1}). The constraint (\ref{divtheta0}) is required to
cancel the tadpole anomalies arising from regularization of Feynman
amplitudes which break the symmetry of the path integral under
diffeomorphisms of the worldsheet $\Sigma$~\cite{CatFel1}.

In the context of open string theory, the Born-Infeld measure
\beq
\Omega_{\rm BI}=\sqrt{\det(g+\mcB)}~\dd^Dx
\label{BImeasure}\eeq
is the canonical choice. Then the divergence-free condition
(\ref{divtheta0}) has the natural physical interpretation as the Born-Infeld
equations of motion (\ref{BIeom}) on the brane. Higher derivative
stringy corrections to the measure (\ref{BImeasure}) can preserve the
cyclicity properties (\ref{Kontcycl1}) and (\ref{Kontcycl2}) of the
deformed product even in the nonassociative cases~\cite{Cycl1,Cycl2}.

\subsection{Poincar\'e-Birkhoff-Witt Theorem\label{PBHTheorem}}

We will now specialize to the case of the worldvolume $M=\real^D$. The
Kontsevich formula in this instance simplifies
drastically~\cite{Kont1,CatFel1}. However, even then a complete
description of it would go beyond the scope of the present article. We
shall therefore develop a ``dual'' description of the Kontsevich
product along the lines of what we did for the Moyal product in
Section~\ref{WeylRep}. Under suitable circumstances, this provides a
more tractable way of extracting the asymptotic expansion coefficients
in concrete calculations.

The basic set-up described above can be described in terms of
noncommuting coordinates which are given by abstract operators $\hat
x^i$ obeying commutation relations of the type
\beq
\bigl[\hat x^i\,,\,\hat x^j\bigr]=\ii\theta^{ij}(\hat x) \ .
\label{hatxthetax}\eeq
Assume that the tensor function $\theta^{ij}(x)$ has a power series
expansion of the form
\beq
\theta^{ij}(x)=\vartheta^{ij}+C^{ij}{}_k~x^k-\ii\bigl(q\,
R^{ij}{}_{kl}-\delta^j{}_k\,\delta^i{}_l\bigr)~x^k\,x^l
+O\big(x^3\big)
\label{thetaxexp}\eeq
with $\vartheta^{ij}$, $C^{ij}{}_k$ and $q\,R^{ij}{}_{kl}$
constants. The terms in this expansion correspond respectively to the
canonical Moyal spaces $[\hat x^i,\hat x^j]=\ii\vartheta^{ij}$ studied
in previous sections, the Lie algebra type noncommutative spaces
$[\hat x^i,\hat x^j]=\ii C^{ij}{}_k~\hat x^k$ associated to the
quantizations of {\it linear} Poisson structures on $\real^D$, and the
quantum (or $q$-deformed) spaces $[\hat x^i,\hat
x^j]=q\,R^{ij}{}_{kl}~\hat x^k\,\hat x^l$. What makes all three of these
particular instances special is that they fulfill the requirements of
the {\it Poincar\'e-Birkhoff-Witt theorem}~\cite{GaugeNC}, and we will
assume that the generic case (\ref{hatxthetax}) also satisfies this
property.

The algebra $\palg_0=\complex(\real^D)$ of polynomial functions on the
vector space $\real^D$ is naturally isomorphic to the symmetric tensor
algebra of the dual vector space $(\real^D)^\vee$. Let
$\palg_\theta=U(\real_\theta^D)$ be the universal enveloping algebra
of the coordinate algebra generated by the operators $\hat x^i$ modulo
the commutation relations (\ref{hatxthetax}). Then the
Poincar\'e-Birkhoff-Witt theorem asserts that the map
\beq
\palg_0~\longrightarrow~\palg_\theta \ , \quad
x^{i_1}\cdots x^{i_n}~\longmapsto~\NO ~\hat x^{i_1}\cdots\hat x^{i_n}~\NO
\label{PBHmap}\eeq
is a linear isomorphism, where $\NO-\NO$ denotes an ordering for elements of
the basis of monomials for $\palg_\theta$. As in Section~\ref{WeylRep} (see
Property~2), we may use this isomorphism to transport the algebraic
structure on the noncommutative algebra $\palg_\theta$ to the algebra
of polynomial functions on $\real^D$ and hence define a star-product
on $\palg_0$. Because the product on the right-hand side of
(\ref{PBHmap}) is taken in the universal enveloping algebra, this
star-product is noncommutative and associative. Since $\palg_0$ is
dense in $\CC^\infty(\real^D)$, this naturally extends to a
star-product defined on Schwartz functions.

While different choices of ordering in (\ref{PBHmap}) lead to
different explicit star-products, we will see in
Section~\ref{Formality} below that they are all equivalent, in the
sense of (\ref{newstarequiv}). The canonical choice is the symmetric
(or Weyl) ordering which assigns to any monomial in $x^i$ the totally
symmetrized monomial in $\hat x^i$. We extend this map to arbitrary
Schwartz functions $f\in\CC^\infty(\real^D)$ by using the same formula
(\ref{MoyalWeylSymbdef}) as in the case of the Moyal product. The
resulting star-products $f\star g$ can be computed by using the
commutation relations (\ref{hatxthetax}) and the
Baker-Campbell-Hausdorff formula~\cite{BehrSyk1}. The leading terms in
a formal asymptotic expansion in powers of $\theta$ coincide with
those of the Kontsevich formula~(\ref{Kontexp}).

An important special instance of this construction is that of a linear
Poisson structure in (\ref{thetaxexp}), representing the commutation
relations of a Lie algebra $\mfg$. Then this procedure yields a
deformation quantization of the Kirillov-Kostant Poisson structure on
the dual $\mfg^\vee$, which coincides with the formal deformation
quantization obtained from the standard coadjoint orbit method. In
this case the star-product constructed here is called the {\it Gutt
  product}~\cite{Gutt1} and it is equivalent to the Kontsevich
product~\cite{Kath1,Shoikhet1,Dito1}. It is identical to the
Kontsevich formula only when
$\mfg$ is a nilpotent Lie algebra~\cite{Kath1}, i.e. the third order
Lie bracket of any four elements of $\mfg$ vanishes. Nilpotent Lie
algebras also have the powerful property that the cyclicity properties
(\ref{Kontcycl1}) and (\ref{Kontcycl2}) hold for the canonical
translationally-invariant flat measure $\Omega=\dd^Dx$ on $\mfg^\vee$
with $D=\dim(\mfg)$.

\subsection{Diffeomorphisms\label{CurvedDiffs}}

We now turn to the implementation of spacetime symmetries on the
noncommutative curved spaces constructed above. If the star-product
originates from a twist element $\mcF\in U(\Vect(\real^D))\otimes
U(\Vect(\real^D))$ as in the previous section, then this is
straightforward to do by using the usual action given by the Lie
derivative
$\mathfrak{L}_{(-)}(-):\Vect(\real^D)\times\CC^\infty(\real^D)\to
\CC^\infty(\real^D)$ and the decompositions
\beq
\mcF~:=~\sum_n\,\mfff^n\otimes\mfff_n \quad \mbox{and} \quad
\mcF^{-1}~:=~\sum_n\,\mff^n\otimes\mff_n \qquad \mbox{with}
\qquad f\star g~=~\sum_n\,\mff^n(f)~\mff_n(g)
\label{twistdecompsum}\eeq
for $f,g\in\CC^\infty(\real^D)$. Then the twisted coproduct
$\Delta_\mcF$ and deformed action of twisted
diffeomorphisms on $\alg_\theta$ are defined respectively
by~\cite{Wessgrav2}
\bea
\Delta_\mcF(X)&=&X\otimes1+\sum_{k,l,m,n}\,\mff^k~\mff^n~\mfff_l
\bigl(\,\mff^m\,S(\mff_m)\,S^{-1}(\mfff_k)\bigr)\otimes\mff_n~
\mfff^l(X) \ , \nonumber\\
X^\star\triangleright f&=&\sum_n\,\mathfrak{L}_{\mff^n(X)}\circ
\mathfrak{L}_{\mff_n}(f)
\label{defdifftwist}\eea
for $X\in\Vect(\real^D)$. It is straightforward to check that this
action is well-defined and compatible with the star-product in
$\alg_\theta$. One can now repeat the construction of
Section~\ref{TwistDiff} to get a noncommutative theory of gravity
which is covariant under deformed diffeomorphisms and is
coordinate-independent~\cite{Wessgrav2}. The three types of
noncommutative spaces appearing in the expansion (\ref{thetaxexp}) all
fall into this category, their twist elements being given by
exponentiating sets of mutually commuting smooth vector fields on
$\real^D$.

We can, however, consider more general deformations by
exploiting the fact that the generic Poisson diffeomorphism group
${\rm Diff}_\theta(\real^D)$ will be far richer now than in the case of
constant $\theta$. Let $\Vect_\theta(\real^D)$ be the Lie algebra of
Poisson vector fields $X$ obeying
\beq
[X,\theta]_{\rm S}=0 \ .
\label{XthetaS0}\eeq
Such vector fields are derivations of the corresponding Poisson
bracket, satisfying the Leibniz rule
\beq
X\big(\{f,g\}_\theta\big)=\big\{X(f)\,,\,g\big\}_\theta+
\big\{f\,,\,X(g)\big\}_\theta
\label{PoissonLeibniz}\eeq
for all $f,g\in\CC^\infty(\real^D)$. We assume that, corresponding to
each Poisson vector field $X$, there exists a polydifferential
operator $\delta_X^\star$ on $\alg_\theta$ which is a derivation of
the star-product,
\beq
\delta_X^\star(f\star g)=(\delta_X^\star f)\star
g+f\star(\delta_X^\star g) \ .
\label{polydiffLeibniz}\eeq
We can write this condition in a global form analogous to
(\ref{XthetaS0}) by introducing the graded Gerstenhaber Lie bracket
between any two polydifferential operators ${\rm D}_1$ and ${\rm
  D}_2$, of degrees $|{\rm D}_1|$ and $|{\rm D}_2|$, through
\beq
[{\rm D}_1,{\rm D}_2]_{\rm G}={\rm D}_1\diamond{\rm D}_2-
(-1)^{(|{\rm D}_1|-1)\,(|{\rm D}_2|-1)}~{\rm D}_2\diamond{\rm D}_1
\label{Gertenbracket}\eeq
where the graded Gerstenhaber product is given by
\beq
{\rm D}_1\diamond{\rm D}_2=
\sum_{l=1}^{n_1}\,(-1)^{(n_2-1)\,(l-1)}~{\rm D}_1\circ\bigl(
\id^{\otimes(l-1)}\otimes{\rm D}_2\otimes\id^{\otimes(n_1-l)}\bigr)
\label{diamondprod}\eeq
acting on $(\alg_\theta)^{\otimes(n_1+n_2-1)}$. Let ${\rm
  B}_\theta$ be the bidifferential operator implementing the
star-product, i.e. $f\star g:={\rm B}_\theta(f,g)$, given by
(\ref{Kontformalexp}). Then (\ref{polydiffLeibniz}) is equivalent to
\beq
[\delta^\star_X,{\rm B}_\theta]_{\rm G}=0 \ .
\label{deltaXBG0}\eeq

The existence of such a map $\delta^\star_{(-)}$ between Poisson
vector fields in $\Vect_\theta(\real^D)$ and derivations of the
star-product will be established in generality in
Section~\ref{Formality} below. It can be constructed as an asymptotic
series in powers of $\theta$ by using the Weyl-ordered star-product of
Section~\ref{PBHTheorem} above along with the expansion
\beq
\delta_X^\star=X+\sum_{n=2}^\infty\,\xi_X^{i_1\cdots i_n}~\partial_{i_1}
\cdots\partial_{i_n} \ .
\label{deltaXexp}\eeq
Expanding (\ref{polydiffLeibniz}) order by order in $\theta$ and
using (\ref{XthetaS0}) gives explicitly~\cite{BehrSyk1}
\beq
\delta_X^\star=X+\mbox{$\frac1{12}$}\,\theta^{lk}\,\partial_k\theta^{im}\,
\partial_l\partial_mX^j\,\partial_i\partial_j-\mbox{$\frac1{24}$}\,
\theta^{lk}\,\theta^{im}\,\partial_l\partial_iX^j\,\partial_k\partial_m
\partial_j+O\big(\theta^3\big) \ .
\label{deltaXexpO3}\eeq
This mapping in fact establishes a one-to-one correspondence. If $\rm
D$ is any derivation of the star-product, then there exists a vector
field $X_{\rm D}\in\Vect_\theta(\real^D)$ such that
\beq
\delta_{X_{\rm D}}^\star={\rm D} \ .
\label{deltaXonetoone}\eeq
In particular, if $X,Y\in\Vect_\theta(\real^D)$ then
$[\delta_X^\star,\delta_Y^\star]_{\rm G}$ is again a derivation of the
star-product and we conclude
\beq
\big[\delta_X^\star\,,\,\delta_Y^\star\big]=
\delta^\star_{[X,Y]_\star} \ ,
\label{deltaXYstar}\eeq
where $[X,Y]_\star$ is a deformation of the Lie bracket
(\ref{VectRDbracket}) of commutative vector fields on $\real^D$. Using
(\ref{deltaXexpO3}) one computes~\cite{BehrSyk1}
\bea
[X,Y]_\star&=&[X,Y]+\mbox{$\frac1{12}$}\,\theta^{lk}\,
\partial_k\theta^{im}\,\bigl(\partial_l\partial_mX^j\,\partial_i\partial_j
Y^p-\partial_l\partial_mY^j\,\partial_i\partial_jX^p\bigr)\,\partial_p
\nonumber \\ && -\,\mbox{$\frac1{24}$}\,\theta^{lk}\,\theta^{im}\,
\bigl(\partial_l\partial_iX^j\,\partial_k\partial_m\partial_jY^p-
\partial_l\partial_iY^j\,\partial_k\partial_m\partial_jX^p\bigr)\,
\partial_p+O\big(\theta^3\big) \ .
\label{XYstarexpO3}\eea

With these ingredients at hand we can now easily formulate gauge
theory on a curved noncommutative space with a Poisson structure
$\theta$ which is compatible with a frame $e_a=e_a^i(x)\,\partial_i$,
$a=1,\dots,D$ in which the given metric tensor $G$ on $\real^D$ is
constant, i.e. $\eta_{ab}=e_a^i\,G_{ij}\,e_b^j$ with
$e_a\in\Vect_\theta(\real^D)$. Given Poisson vector fields $X,Y$ and a
$\UU(1)$ gauge connection $A=A_i(x)~\dd x^i$, we define the
covariant derivative of a scalar field $f$ by
\beq
D_Xf=\delta_X^\star f-\ii A(X)\star f
\label{DXfdeltadef}\eeq
and the corresponding field strength as
\beq
F(X,Y)=-\ii[D_X,D_Y]_{\star}+\ii D_{[X,Y]_\star} \ .
\label{FXYstrengthdef}\eeq
The properties of the maps $\delta_{(-)}^\star$ and $[-,-]_\star$
ensure that (\ref{FXYstrengthdef}) is a function in
$\CC^\infty(\real^D)$ and not a polydifferential operator. We can now
evaluate the noncommutative field strength (\ref{FXYstrengthdef}) on
the frame $e_a$ and
define
\beq
F_{ab}=F(e_a,e_b) \ .
\label{Fabdef}\eeq
Picking a measure $\Omega$ on $\real^D$ with the properties
(\ref{divtheta0})--(\ref{Kontcycl2}), the action for $\UU(1)$
noncommutative Yang-Mills theory on the curved background is defined
as in the case of flat space and is given by~\cite{BehrSyk1}
\beq
{\mathcal S}_{\rm NCYM}:=\int\,\Omega~\eta^{ab}\,\eta^{cd}\,F_{ac}\,
F_{bd} \ .
\label{SNCYMcurveddef}\eeq
In the commutative limit $\theta\to0$ it reduces to the usual
Yang-Mills action on $\real^D$ with a curved metric $G$, provided that
one chooses $\Omega\big|_{\theta=0}=\sqrt{\det(G)}~\dd^Dx$ to be the
corresponding riemannian volume form. The crux of this construction is
the possibility to find Poisson structures and compatible frames,
which is not always an easy task for complicated star-products
(see~\cite{CerFrame1} for an investigation on certain quantum
spaces). We will see some explicit examples in the next section.

\subsection{Formality Theorem\label{Formality}}

The {\it formality theorem}~\cite{Kont1} is at the very heart of the program of
global deformation quantization of the algebras of functions on arbitrary
Poisson manifolds. It states that the differential graded Lie algebra
of polydifferential operators, equiped with the Gerstenhaber bracket,
is ${\rm L}_\infty$-quasiisomorphic to its cohomology, given by the
differential graded Lie algebra of polyvector fields equiped with the
Schouten-Nijenhuis bracket. This result has several important uses. It
leads to closed expressions for certain star-products which are
equivalent to the Kontsevich product, and it is useful for addressing
questions of existence and the relationships between Poisson bivector
fields and star-products. For example, it can be used to give a closed
form for the map $\delta_{(-)}^\star$ introduced above, and also to
formally assert gauge equivalences between different
star-products. Moreover, the formality formulas enable one to trace
the generic nonassociativity of a star-product to the
Schouten-Nijenhuis bracket $[\theta,\theta]_{\rm S}$, which in the
open string setting is proportional to the NS--NS field strength
(\ref{HddB}). Its drawback is that it is not a particularly useful
tool for explicit concrete calculations.

The {\it formality map} is a collection of skew-symmetric multilinear maps
$\mathcal{U}_n$, $n\in\nat_0$, that take $n$ polyvector fields to an
$m$-differential operator and fulfill a combinatorial recursion
relation known as the {\it formality condition}. If $X_1,\dots,X_n$ are
polyvector fields of gradings (degrees) $k_1,\dots,k_n$, then ${\mathcal
  U}_n(X_1,\dots,X_n)$ is a polydifferential operator of grading
(degree)
\beq
m=2-2n+\sum_{i=1}^n\,k_i \ .
\label{polydiffdegreem}\eeq
In particular, the first-order term $\mcU_1$ coincides with the
Hochschild-Kostant-Rosenberg map which takes a $k$-vector field to a
$k$-differential operator defined by
\beq
\mcU_1\big(X^{i_1\cdots i_k}\,\partial_{i_1}\wedge\cdots\wedge
\partial_{i_k}\big)=X^{i_1\cdots
  i_k}\,\mu_0\circ\big(\partial_{i_1}\otimes\cdots\otimes\partial_{i_k}
\big) \ ,
\label{mcU1def}\eeq
where here $\mu_0(f_1\otimes\cdots\otimes f_k)=f_1\cdots f_k$ is the
pointwise product on $(\alg_0)^{\otimes k}$. The collection of
formality maps $(\mcU_n)_{n\geq0}$ then satisfy the formality
conditions
\beq
\sum_{n_1+n_2=n}\,\mcQ_2\circ\bigl(\mcU_{n_1}\otimes\mcU_{n_2}
\bigr)~=~\sum_{l=0}^{n-2}\,\mcU_{n-1}\circ\bigl(\id^{\otimes l}
\otimes\mcQ_2\otimes\id^{\otimes(n-l-2)}\bigr) \qquad \mbox{for}
\quad n\geq1
\label{formalitycond}\eeq
on the space of symmetric tensors over the algebra
$\CC^\infty(\real^D)$, where the quadratic form $\mcQ_2$ is defined by
$\mcQ_2(\DD_1,\DD_2)=(-1)^{(|\DD_1|-1)\,|\DD_2|}\,[\DD_1,\DD_2]_{\rm
  G}$ on polydifferential operators and by
$\mcQ_2(X,Y)=-(-1)^{(k_X-1)\,k_Y}\,[X,Y]_{\rm S}$ on polyvector fields.

Given an arbitrary bivector field $\theta$, we define a star-product
through the bidifferential operator $\BB_\theta$ given by
\beq
f\star g~=~\BB_\theta(f,g)~:=~\sum_{n=0}^\infty\,\frac1{n!}\,
\mathcal{U}_n(\theta,\dots,\theta)(f,g) \ .
\label{fstargformality}\eeq
We also introduce special polydifferential operators
\bea
\psi_1(X)&=&\sum_{n=1}^\infty\,\frac1{(n-1)!}\,\mcU_n(X,\theta,\dots,
\theta) \ , \nonumber\\ 
\psi_2(X,Y)&=&\sum_{n=2}^\infty\,\frac1{(n-2)!}\,\mcU_n(X,Y,\theta,\dots,
\theta) \ .
\label{psispecial}\eea
If $f\in\CC^\infty(\real^D)$ and $X,Y\in\Vect(\real^D)$, then also
$\psi_1(f),\psi_2(X,Y)\in\CC^\infty(\real^D)$. For
$X\in\Vect(\real^D)$ introduce the one-differential operator
\beq
\delta_X^\star=\psi_1(X) \ .
\label{deltaXPhiX}\eeq
Then the formality conditions (\ref{formalitycond}) lead to the
Gerstenhaber brackets
\bea
\bigl[\BB_\theta\,,\,\BB_\theta\bigr]_{\rm G}&=&
\psi_1\big([\theta,\theta]_{\rm S}\big) \ , \nonumber\\
\bigl[\delta_X^\star\,,\,\BB_\theta\bigr]_{\rm G}&=&
\psi_1\bigl([X,\theta]_{\rm S}\bigr) \ , \nonumber\\
\bigl[\delta_X^\star\,,\,\delta_Y^\star\bigr]_{\rm G}+
\bigl[\psi_2(X,Y)\,,\,\BB_\theta\bigr]_{\rm G}&=&
\delta_{[X,Y]_{\rm S}}^\star+\psi_2\bigl([\theta,Y]_{\rm S}\,,\,X
\bigr)-\psi_2\bigl([\theta,X]_{\rm S}\,,\,Y\bigr) \ .
\label{Gerstbrackets}\eea
The first condition in (\ref{Gerstbrackets}) measures the failure of
associativity of the star-product (\ref{fstargformality}), the second
condition gives the failure of the operators (\ref{deltaXPhiX}) in
producing derivations of the star-product (\ref{fstargformality}), and
the last condition measures the failure of (\ref{deltaXPhiX}) in
giving a representation of the Lie algebra $\Vect(\real^D)$.

Suppose now that $\theta$ is a Poisson bivector field, with vanishing
Schouten-Nijenhuis bracket (\ref{thetaSchouten}), and that
$X,Y\in\Vect_\theta(\real^D)$ are Poisson vector fields, obeying
(\ref{XthetaS0}). Then the relations (\ref{Gerstbrackets}) evaluated
on functions $f,g,h\in\CC^\infty(\real^D)$ respectively become
\bea
f\star(g\star h)&=&(f\star g)\star h \ , \nonumber \\
\delta_X^\star(f\star g)&=&(\delta_X^\star f)\star g+f
\star(\delta_X^\star g) \ , \nonumber \\
\bigl(\bigl[\delta_X^\star\,,\,\delta_Y^\star\bigr]-
\delta_{[X,Y]}^\star\bigr)(f)&=&\bigl[\psi_2(X,Y)\,,\,f
\bigr]_\star \ .
\label{GerstbracketsPoisson}\eea
In particular, from the last equation in (\ref{GerstbracketsPoisson})
we see that the map $\delta_{(-)}^\star$ preserves the Lie bracket of
vector fields up to an inner automorphism. We may cast this equation
into the form (\ref{deltaXYstar}) with the deformed Lie bracket given
explicitly by
\beq
\big[X\,,\,Y\big]_\star=\big[X\,,\,Y
\big]+\bigl[\theta\,,\,\psi_1^{-1}\circ
\psi^{\phantom{1}}_2(X,Y)\bigr]_{\rm S} \ .
\label{defbracketexpl}\eeq

\subsection{A-Branes\label{ABranes}}

Under suitable conditions, the quantization of D-branes in the Poisson
sigma-model of Section~\ref{KontFormula} above may be consistently
carried out. When the branes wrap coisotropic submanifolds,
i.e. worldvolumes $W$ defined by first-class constraints, then they
play the role of D-branes for the open topological A-model string
theory (called {\it A-branes} for short). In this case the
quantization can be related to the deformation quantization in the
induced Poisson brackets~\cite{CatFel2,CatFel3}, as we now describe
explicitly. Branes defined by second-class constraints may also be
treated by quantizing Dirac brackets on the
worldvolumes~\cite{Falceto1}.

Let $\iota:W\hookrightarrow\real^D$ be the worldvolume embedding of a
D-brane, given by first-class constraints $f_a=0$ with
$f_a\in\CC^\infty(\real^D)$. This means that the functions $f_a$
Poisson commute with every function on $\real^D$,
i.e. $\theta(f_a,f)=0\quad\forall f\in\CC^\infty(\real^D)$. Let
$\pi:\real^D\to W$ be the canonical projection. Let
$\iota^*:\CC^\infty(\real^D)\to\CC^\infty(W)$ and
$\pi^*:\CC^\infty(W)\to\CC^\infty(\real^D)$ be the corresponding
pullbacks. Assume that $f_a$ star-commute with every function in
$\CC^\infty(\real^D)$. Then the star-product $\star$ on $\real^D$ can
be consistently restricted to a star-product $\star_0$ on the
worldvolume $W$ defined by
\beq
f_0\star_0g_0=\pi^*(f_0)\star\pi^*(g_0)
\label{starWdef}\eeq
for $f_0,g_0\in\CC^\infty(W)$. There is a compatibility condition
\beq
\iota^*(f\star g)=\iota^*(f)\star_0\iota^*(g)
\label{compcondiota}\eeq
for $f,g\in\alg_0=\CC^\infty(\real^D)$, and one has an isomorphism
$\CC^\infty(W)=\alg_0/\mathcal{I}$ where $\mathcal{I}$ is the
two-sided ideal of the algebra $\alg_0$ generated by the Casimir
constraints $f_a$.

This construction is a noncommutative version of Poisson
reduction~\cite{Waldmann1}, with the Poisson ideal $\mcI$ implementing
the geometric requirement that the Seiberg-Witten bivector field
$\theta$ be tangent to the worldvolume $W$. With the above conditions
fulfilled, one can also consistently define the actions of twisted
spacetime symmetries on $W$ with
\bea
\Delta_W(X_0)\triangleright(f_0\otimes g_0)&=&(\iota^*\otimes\iota^*)
\circ\Delta_\theta(X)\triangleright
\bigl(\pi^*(f_0)\otimes\pi^*(g_0)\bigr) \
, \nonumber \\
X_0^{\star_0}\triangleright f_0&=&\iota^*\circ X^\star\triangleright\bigl(
\pi^*(f_0)\bigr)
\label{redsymmetries}\eea
for $X\in\Vect(\real^D)$ and $f_0,g_0\in\CC^\infty(W)$. However, in
some cases not all of the above criteria are met. In such instances
a relative version of the formality theorem of Section~\ref{Formality}
above is available for obtaining explicit quantizations of
D-submanifolds of a noncommutative spacetime~\cite{CatFel3}. One has
the natural notions of {\it relative} polyvector fields on
$(\real^D,W)$, which form a differential graded Lie algebra with the induced
Schouten-Nijenhuis bracket, and of {\it relative} polydifferential
operators, which also form a differential graded Lie algebra with
respect to the induced Gernstenhaber bracket. Then similarly to
Section~\ref{Formality} above, one constructs an ${\rm
  L}_\infty$-quasiisomorphism $(\mcU_n)_{n\geq0}$ between the
differential graded Lie algebras of relative polyvector fields and of
relative polydifferential operators. This result implies that there is
a duality between A-branes and supersymmetric spacetime-filling
D-branes in the Poisson sigma-model. The perturbative
expansion of the sigma-model path integral around the corresponding
non-trivial classical solutions leads to a generalization of the
Fukaya ${\rm A}_\infty$-category of topological D-branes. Other
aspects of noncommutative string theory in curved backgrounds can be
found in~\cite{Kar1,Kar2}.

\newsection{Superstring Backgrounds\label{SuperBacks}}

In this final section we will describe some examples of curved
spacetimes to illustrate the general formalism of the past two
sections. We shall study the noncommutative gauge theories of
various classes of D-branes in certain tractable curved supergravity
backgrounds of Type~II superstring theory. We will emphasize both
algebraic and geometric features of the spacetime symmetries in these
instances.

\subsection{$\AdS_3\times\S^3$\label{S3DBranes}}

Consider the exact supergravity background
$M=\AdS_3\times\S^3\times\mathcal{M}_4$ of ten-dimensional string
theory, where $\mathcal{M}_4$ is any exact four-dimensional background
such as flat space or a K3-surface. Without the $\mathcal{M}_4$ factor the
background is a vacuum solution of the minimal chiral supergravity in
six dimensions. We are interested in the class of symmetric D-branes in
this spacetime which wrap two-spheres $\S^2\subset\S^3$. This is
the simplest and best understood example of curved D-branes, and we
will only use it to highlight issues related to the twisted spacetime
symmetries of Section~\ref{TwistSym} and to the associativity
properties of Section~\ref{NCGTCurved} (see~\cite{Schomrev} for a more
general in-depth review of symmetric D-branes in curved
backgrounds). In the case at hand these two features merge to give a
nice illustration of the manner in which deformations lead to quantum
group symmetries of systems of D-branes.

From an algebraic perspective, the D-branes in question wrap
conjugacy classes of the Lie group $\SU(2)\cong\S^3$~\cite{ARS1}. Let
$k\in\nat\cup\{\infty\}$. Then the dynamics of open strings ending on
such a D-brane is described by a particular worldsheet boundary conformal field
theory, the $\SU(2)$ Wess-Zumino-Witten model at level $k$. The
radius of the sphere $\S^3$ is given by $R=\sqrt{k/2\pi}$, and the
NS--NS field strength is $H=\frac1k~\Omega_{\S^3}$ with
$\Omega_{\S^3}$ the standard round volume form on $\S^3$. In the
boundary conformal field theory, there are $k+1$ boundary conditions
labelled by $N=0,1,\dots,k$, and primary fields represented by
boundary vertex operators $Y_i^I$ with $I=0,1,\dots,\min(N,k-N)$ and
$i=1,\dots,2I+1$. The corresponding operator product expansion on zero
modes of the open string embedding fields gives
an abstract algebra $\alg_k$ generated by the $Y_i^I$ with the
product~\cite{ARS1}
\beq
Y_i^I\star Y_j^J~=~\mu_k\big(Y_i^I\otimes Y_j^J\big)~:=~
\sum_{L,l}\,\Bigl[{}^I_i~{}^J_j~{}^L_l\Bigr]\,\Bigl\{
{}^I_N~{}^J_N~{}^L_N\Bigr\}_q~Y_l^L \ ,
\label{Ystar}\eeq
where the square brackets denote classical Clebsch-Gordan coefficients
and the curly brackets are $q$-deformed $6j$-symbols for $\SU(2)$ with
\beq
q=\e^{\pi\ii/(k+2)} \ .
\label{qkdef}\eeq
This defines a finite-dimensional quasi-associative algebra $\alg_k$
which is covariant under the natural action of the $\mfs\mfu(2)$ Lie
algebra.

Let us first consider the semi-classical limit $k\to\infty$, whereby
the dynamics of this system simplifies drastically. In this limit the
radius $R\to\infty$, so that the sphere $\S^3$ grows and approaches
flat space $\real^3$, while $H\to0$, so that one can anticipate an
associative noncommutative worldvolume gauge theory from the general
considerations of the previous section. Furthermore, $q\to1$ and the
quantum $6j$-symbols in (\ref{Ystar}) become ordinary $6j$-symbols of
$\SU(2)$. In this case (\ref{Ystar}) describes an associative
algebra $\alg_\infty$ which coincides with the classic {\it fuzzy
  sphere} $\S_N^2$~\cite{Madore1,GKP1}. Let $\ell_i$, $i=1,2,3$ be the
generators of the irreducible spin $\frac N2$ representation of
$\mfs\mfu(2)$ obeying the relations
\beq
[\ell_i,\ell_j]~=~\ii\epsilon_{ij}{}^k~\ell_k \qquad \mbox{and}
\qquad \ell_1^2+\ell_2^2+\ell_3^2~=~\mbox{$\frac N2\,\big(\frac N2+1
\big)$}~=:~\Lambda_N^{-2}
\label{ellirels}\eeq
in $U(\mfs\mfu(2))$. Then the coordinate generators
$x_i:=Y_i^1=\Lambda_N\,\ell_i$ of $\S_N^2$ satisfy
\beq
\epsilon^{ij}{}_k~x_i\,x_j~=~\Lambda_N\,x_k \qquad \mbox{and}
\qquad x_1^2+x_2^2+x_3^2~=~1 \ .
\label{xirels}\eeq
This gives the standard Kirillov-Kostant symplectic structure on
the quantized coadjoint orbits $\SU(2)/\UU(1)\cong\S^2$ of the Lie
algebra $\mfs\mfu(2)\cong\real^3$. Since $\Lambda_N\to0$ in the limit
$N\to\infty$, the algebra $\S_\infty^2$ coincides with the algebra of
functions on the standard unit sphere
$x_i:\S^2\hookrightarrow\real^3$.

The isometry group of rotations of the sphere yields a natural adjoint
action of $\mfs\mfu(2)$ on $\S_N^2$ given by
\beq
\ell_i\triangleright x_j~:=~{\rm ad}_{\ell_i}(x_j)~=~
[\ell_i,x_j]~=~\ii\epsilon_{ij}{}^k~x_k \ .
\label{su2actionSN2}\eeq
This in turn leads to a decomposition of the $\complex$-algebra of
polynomial functions of the $x_i$ (as in Section~\ref{PBHTheorem})
given by
\beq
\S_N^2=\underline{1}\,\oplus\,\underline{3}\,
\oplus\,\cdots\,\oplus\,\underline{2N+1} \ ,
\label{SN2su2decomp}\eeq
where generally $\underline{d}$ denotes the irreducible representation of
$\mfs\mfu(2)$ of dimension $d\in\nat$. This decomposition simply
reflects the standard decomposition of homogeneous polynomial
functions on the sphere into spherical harmonics, except that now the
maximum allowed angular momentum is $\frac N2$. It also identifies the
fuzzy sphere as a full matrix algebra
$\S_N^2\cong\mat_{N+1}$. One thereby obtains a
finite-dimensional algebra of functions on the sphere.

Let us now consider the generic stringy regime in which $k<\infty$. In
this case one can trade the nonassociativity of the algebra
(\ref{Ystar}) for a $q$-deformation of the $\SU(2)$ symmetry group by
using the standard Drin'feld twist element $\mcF\in
U(\mfs\mfu(2))\otimes U(\mfs\mfu(2))$ to define an {\it associative}
product for $f,f'\in\alg_k$ by the usual twisted product
\beq
f~\widetilde{\star}~f':=\mu_k\circ\mcF^{-1}\triangleright\big(f\otimes
f'\big) \ .
\label{tildestardef}\eeq
The algebra relations (\ref{Ystar}) are then modified to
\beq
Y_i^I~\widetilde{\star}~Y_j^J=
\sum_{L,l}\,\Bigl[{}^I_i~{}^J_j~{}^L_{l'}\Bigr]_q\,\Bigl\{
{}^I_N~{}^J_N~{}^L_N\Bigr\}_q\,G_{(q)}^{l'l}~Y_l^L \ ,
\label{Ystartilde}\eeq
where now the square brackets denote quantum Clebsch-Gordan
coefficients of $\SU(2)$ and $G_{(q)}$ is the $q$-deformed flat metric of
$\real^3$ given by
\beq
\big(G_{(q)}^{ij}\big)=\begin{pmatrix} & & \scriptstyle q^{-1} \\
 & \scriptstyle \id_3 & \\ \scriptstyle q & & \end{pmatrix} \ .
\label{qdefmetric}\eeq
This defines a finite-dimensional {\it associative} algebra $\alg_k$
which has the structure of a quantum fuzzy sphere
$\S_{q,N}^2$~\cite{GMStein1}, defined by coordinate generators
$x_i:=Y_i^1$ obeying the relations
\beq
\epsilon_{(q)}^{ij}{}_k~x_i\,x_j~=~\Lambda_N^{(q)}\,x_k \qquad
\mbox{and} \qquad x_i\,G_{(q)}^{ij}\,x_j~=~1 \ .
\label{qsphererels}\eeq
Here $\Lambda_N^{(q)}:=[2]_q/\sqrt{[N]_q\,[N+2]_q}$ and
$[x]_q=\frac{q^x-q^{-x}}{q-q^{-1}}$ is the $q$-number associated to
$x\in\real$ with $[x]_q\to x$ in the limit $q\to1$, while the symbols
$\epsilon_{(q)}^{ij}{}_k$ are given in terms of quantum
Clebsch-Gordan coefficients.

Under the twisting defined above, the natural $\mfs\mfu(2)$ rotational
symmetry of the fuzzy sphere $\S_N^2$ is deformed to a covariant
action of the noncocommutative Hopf algebra $U_q(\mfs\mfu(2))$ (the
quantum universal enveloping algebra of $\mfs\mfu(2)$ or quantum
$\SU(2)$ group), which as an algebra is generated by elements
$K,E^\pm$ modulo the commutation relations
\beq
\big[K\,,\,E^\pm\big]~=~\pm\,2E^\pm \qquad \mbox{and} \qquad
\big[E^+\,,\,E^-\big]~=~[K]_q \ .
\label{suq2commrels}\eeq
The $x_i$ in (\ref{qsphererels}) may then be expressed in terms of the
irreducible spin $\frac N2$ representation of $U_q(\mfs\mfu(2))$. The
action of a generic element $X\in U_q(\mfs\mfu(2))$ on the algebra
$\alg_k$ is given by
\beq
X\triangleright x_i=x_j\,\Pi^j{}_i(X)
\label{Uqsu2genaction}\eeq
where $\Pi$ denotes the spin~$1$ representation of
$U_q(\mfs\mfu(2))$. This action leads to the decomposition
\beq
\S_{q,N}^2=\underline{1}\,\oplus\,\underline{3}\,
\oplus\,\cdots\,\oplus\,\underline{2N+1}
\label{SqN2su2decomp}\eeq
which identifies $\S_{q,N}^2\cong\mat_{N+1}$. Thus the $\alg_k$
for all $k\leq\infty$ all have the same underlying {\it algebra} as the
fuzzy sphere $\S^2_N$, but a different coalgebra structure.

The significance of this quantum group symmetry manifests itself most
profoundly in the corresponding noncommutative gauge theory of the
symmetric D-branes wrapping $\S^2$~\cite{GMStein1}. For $q\neq1$
gauge transformations are realized by a quotient of
$U_q(\mfs\mfu(2))$, giving the algebra $\alg_k$ as a quantum
homogeneous space, with a non-trivial coproduct. This leads to a new
kind of gauge symmetry intimately tied to the noncommutative spacetime
symmetries along the lines of what we described in
Section~\ref{SymCNGT}. In the semi-classical regime, when one
considers the noncommutative foliation
$\real_\infty^3:=\bigcup_{N\geq1}\,\S^2_N$ of $\real^3$ by
noncommutative D2-branes (i.e. all fuzzy spheres taken
together), the cocommutative $U(\mfs\mfu(2))$ isometry algebra is
enhanced to the quantum double ${\sf
  D}(U(\mfs\mfu(2)))=\complex\SU(2)\rtimes U(\mfs\mfu(2))$ with the
coadjoint action on the group algebra
$\complex\SU(2)$~\cite{BatMajid1}. The algebra $\real_\infty^3$ is
covariant under the adjoint action of this quantum group.

\subsection{Melvin Universe\label{MelvinTwist}} 

For the remainder of this paper we will focus our attention on
noncommutative gauge theories in time-dependent backgrounds, which
have potentially important applications to string
cosmology~\cite{HashSethi1,Cai2}. A somewhat tractable class of examples is
provided by the Melvin universe and its
generalizations~\cite{HashThom1,HashThom2,CaiOhta1,ASY1}. The Melvin
universe is a non-asymptotically flat solution of Type~IIB
supergravity which has topology $\real^{1,3}\times\real^6$ and is
supported by the flux of an NS--NS $B$-field. It can be constructed
via a sequence of twists and dualities of flat ten-dimensional
spacetime $\real^{1,2}\times\S^1\times\real^6$ with metric
\beq
\dd s^2=-\dd t^2+\dd r^2+r^2~\dd\phi^2+\dd\zeta^2+\dd\my^2 \ ,
\label{flat10Dmetric}\eeq
where $\my\in\real^6$ and the coordinate $\zeta$ is compactified on a
circle $\S^1$ of radius $R$. This is a flat background of Type~IIB
supergravity. Perform a T-duality transformation of
this circle to get a new circle of radius $\tilde R=1/2\pi\,R$ with
coordinate $\tilde\zeta$. The resulting geometry has an isometry
generated by the vector field $\frac\partial{\partial\phi}$. Using
this isometry we can ``twist'' the compactification by changing the
Killing vector field $\frac\partial{\partial\tilde\zeta}$ along the
compactified direction to
$\frac\partial{\partial\tilde\zeta}+\vartheta~
\frac\partial{\partial\phi}$, where $\vartheta\in\real$ is a
constant. Equivalently, one can replace the line element $\dd\phi$ by
$\dd\phi+\vartheta~\dd\tilde\zeta$. By performing a T-duality
transformation back along $\tilde\zeta$, one obtains in this way the
Melvin universe solution of Type~IIB supergravity given
by~\cite{HashThom1}
\bea
\dd s^2&=&-\dd t^2+\dd r^2+\frac{r^2}{1+\vartheta^2\,r^2}~\dd\phi^2+
\frac1{1+\vartheta^2\,r^2}~\dd\zeta^2+\dd\my^2 \ , \nonumber \\
B&=&\frac{\vartheta\,r^2}{1+\vartheta^2\,r^2}~\dd\phi\wedge\dd\zeta \ .
\label{Melvinuniverse}\eea

Now embed a D3-brane into this background by extending it along
the $(t,r,\phi,\zeta)$ directions and localizing it in the $\my$
directions. Let $\mx=(x^1,x^2)$ be cartesian coordinates in the
$(r,\phi)$-plane. Applying the Seiberg-Witten formula (\ref{GthetagB})
to the closed string background (\ref{Melvinuniverse}) gives the
corresponding open string metric $G$ and noncommutativity bivector
field $\theta$ on the brane in the decoupling limit $\vartheta\to0$ as
\bea
G&=&-\dd t\otimes \dd t+\dd\mx^\top\otimes\dd\mx+\dd\zeta
\otimes\dd\zeta \ , \nonumber \\
\theta&=&\vartheta\,\epsilon^{ij}\,x_i~\mbox{$\frac\partial{\partial
    x^j}\wedge\frac\partial{\partial\zeta}$} \ .
\label{GthetaMelvin}\eea
One easily verifies the Jacobi identity (\ref{thetaPoissoncond}) in
this case and hence the bivector field $\theta$ defines a Poisson
structure, as necessary for an associative star-product. We can
therefore proceed with the general constructions of
Section~\ref{NCGTCurved} in the case of flat space $\real^{1,3}$.

The Poisson structure $\theta$ in (\ref{GthetaMelvin}) is linear, and
the corresponding Poisson brackets give a representation of the
euclidean algebra $\mathfrak{iso}(2)=\mathfrak{so}(2)\rtimes\real^2$
in two dimensions. The quantization of the geometry can thereby be
achieved by computing the corresponding Gutt product described in
Section~\ref{PBHTheorem}. We choose the ordering in
(\ref{PBHmap}) given by placing the rotation generator
$\hat\zeta$ to the far right in any monomial in
$U(\mathfrak{iso}(2))$. Choose a complex structure and regard
$\real^2$ as $\complex$ with holomorphic coordinate $z=x_1+\ii x_2$. The
generators of $\mathfrak{iso}(2)$ act on
$(z,\overline{z}\,)\in\complex$ by the affine transformations
$\e^{\alpha\,\hat\zeta}(z,\overline{z}\,)=
(\e^{\ii\vartheta\,\alpha}\,z,\e^{-\ii\vartheta\,\alpha}\,\overline{z}\,)$
and $\e^{w\,\hat z+\overline{w}\,\hat z^\dag}(z,\overline{z}\,)=
(z+\vartheta\,\overline{w}\,,\,\overline{z}+\vartheta\,w)$. From this
action one can easily read off the group multiplication laws, and then
compute the corresponding star-products using the ordered symbols
(\ref{MoyalWeylSymbdef}). One finds that the Gutt product $*$ in this
case is a twisted product determined by the twist
element~\cite{HallSz1}
\beq
\mcF_*=\exp\left[\,\overline{z}\,
\bigl(\e^{\ii\vartheta\,\partial_\zeta}-1
\bigr)\otimes\partial+z\,\bigl(\e^{-\ii\vartheta\,\partial_\zeta}-1
\bigr)\otimes\overline{\partial}\,\right] \ ,
\label{twistMelvin}\eeq
where $\partial:=\frac\partial{\partial z}$ and
$\overline{\partial}:=\frac\partial{\partial\overline{z}}$. This
star-product is not the same as the canonical Weyl-ordered
star-product~$\star$. However, by the formality theorem the two
star-products are cohomologically equivalent, in the sense that there
exists an algebra isomorphism taking one star-product into the other
as in (\ref{newstarequiv}). The invertible differential operator
$\mathcal{D}=\mathcal{D}_*$ in this case may be computed from the
Baker-Campbell-Hausdorff formula which yields~\cite{HallSz1}
\beq
\mathcal{D}_*=\exp\left[\,\overline{z}\,\partial\left(\mbox{$
\frac{\ii\vartheta\,\partial_{\zeta}}{\e^{\ii\vartheta\,\partial_\zeta}-1}$}-1
\right)-z\,\overline{\partial}\left(\mbox{$
\frac{\ii\vartheta\,\partial_{\zeta}}{\e^{-\ii\vartheta\,\partial_\zeta}-1}$}+1
\right)\right] \ ,
\label{diffopMelvin}\eeq
with inverse given by
\beq
\mcD_*^{-1}=\exp\left[\,\overline{z}\,\partial\left(\mbox{$
\frac{\e^{\ii\vartheta\,\partial_\zeta}-1}{\ii\vartheta\,\partial_{\zeta}}$}-1
\right)-z\,\overline{\partial}\left(\mbox{$
\frac{\e^{-\ii\vartheta\,\partial_\zeta}-1}{\ii\vartheta\,\partial_{\zeta}}$}+1
\right)\right] \ .
\label{diffopMelvininv}\eeq

To write down the action of noncommutative gauge theory in the Melvin
universe, we first need to find a local frame as described in
Section~\ref{CurvedDiffs}~\cite{HashThom2}. First observe that the
Poisson bivector field in (\ref{GthetaMelvin}) takes on the simple
constant form
\beq
\theta=\vartheta~\mbox{$\frac\partial{\partial\phi}\wedge\frac\partial
{\partial\zeta}$}
\label{thetaMelvinpolar}\eeq
in polar coordinates $(r,\phi)$. In these coordinates we may therefore
write down the standard Moyal product $\star_0$ on the algebra of
functions. This star-product is not related to the desired
star-product $\star$ corresponding to the curved background
(\ref{GthetaMelvin}) by any simple change of coordinates. However,
again Kontsevich's formality theorem asserts that the star-products
corresponding to (\ref{GthetaMelvin}) and (\ref{thetaMelvinpolar}) are
equivalent up to the given coordinate transformation. To leading
orders the invertible differential operator $\mathcal{D}=\mathcal{D}_{\star_0}$
implementing the equivalence (\ref{newstarequiv}) is given
by~\cite{HashThom2}
\beq
\mathcal{D}_{\star_0}=1+\mbox{$\frac1{24}$}\,\vartheta^2\,r\,\frac\partial
{\partial r}\frac{\partial^2}{\partial\zeta^2}+O\big(\vartheta^3
\big) \ .
\label{calRdefMelvin}\eeq

In polar coordinates, there is a set of pseudo-orthonormal vector
fields $e_a=e_a^i~\partial_i$, $a=1,2,3,4$ given by
\beq
e_1~=~\mbox{$\frac\partial{\partial t}$} \ , \quad
e_2~=~\mbox{$\frac\partial{\partial r}$} \ , \quad
e_3~=~\mbox{$\frac1r~\frac\partial{\partial\phi}$} \qquad
\mbox{and} \qquad e_4~=~\mbox{$\frac\partial{\partial\zeta}$}
\label{Melvinframe}\eeq
which can be used to define a natural local frame compatible with the
Poisson structure (\ref{thetaMelvinpolar}). These frame fields
are evidently derivations of the Moyal product,
\beq
e_a(f\star_0 g)=(e_af)\star_0 g+f\star_0(e_ag) \ .
\label{eaderivMoyal}\eeq
Using the algebra isomorphism (\ref{calRdefMelvin}) we may now define
the polydifferential operators
\beq
\delta_{e_a}^\star f:=\mcD^{\phantom{1}}_{\star_0}\,e_a\,
\mcD_{\star_0}^{-1}f \ .
\label{deltaeaMelvin}\eeq
From (\ref{eaderivMoyal}) and (\ref{newstarequiv}) it follows that
these operators are derivations of the Kontsevich product, since
\beq
\delta_{e_a}^\star(f\star g)~=~\delta^\star_{e_a}
\mcD_{\star_0}^{\phantom{1}}\bigl((\mcD_{\star_0}^{-1}f)
\star_0(\mcD_{\star_0}^{-1}g)\bigr)~=~(\delta^\star_{e_a}f)\star
g+f\star(\delta^\star_{e_a}g) \ .
\label{deltaeaderivMelvin}\eeq

Noncommutative gauge theory of D3-branes in the Melvin background may
now be defined exactly as prescribed in
(\ref{DXfdeltadef})--(\ref{SNCYMcurveddef}). The final point to
address is the appropriate choice of measure $\Omega$. The
divergence-free conditions (\ref{divtheta0}) in the case at hand read
\beq
\partial_\zeta\Omega~=~0 \qquad \mbox{and} \qquad
z\,\partial\Omega~=~\overline{z}\,\overline{\partial}
\Omega-2\Omega \ .
\label{divtheta0Melvin}\eeq
There are many solutions $\Omega$ to these equations. The most
natural one from a physical perspective is the Born-Infeld measure
(\ref{BImeasure}). However, even before considering the dynamics of the
brane we can find a natural geometric measure $\Omega$ as
follows. The noncommutative frame fields (\ref{deltaeaMelvin}) can be
written out explicitly in the form~\cite{BehrSyk1,HallSz1,Moller1}
\bea
\delta_{e_1}^\star f&=&\partial_tf \ , \nonumber \\
\delta_e^\star f&=&-|z|\star\left(\mbox{$\frac{1-\e^{\ii\vartheta\,
\partial_\zeta}}{\ii\vartheta\,\partial_\zeta}$}\right)\partial f \ ,
\nonumber \\ \delta_{\overline{e}}^\star f&=&|z|\star\left(
\mbox{$\frac{1-\e^{-\ii\vartheta\,\partial_\zeta}}{\ii\vartheta\,
\partial_\zeta}$}\right)\overline{\partial}f \ , \nonumber \\
\delta_{e_4}^\star f&=&\partial_\zeta f \ .
\label{NCframeMelvin}\eea
The Leibniz rule (\ref{deltaeaderivMelvin}) may be checked directly
by using the identities
\beq
z\star f~=~\bigl(\e^{\ii\vartheta\,\partial_\zeta}f\bigr)\star z
\qquad \mbox{and} \qquad \overline{z}\star f~=~
\bigl(\e^{-\ii\vartheta\,\partial_\zeta}f\bigr)\star\overline{z} \ .
\label{zstarfids}\eeq
In the commutative limit $\vartheta\to0$, these polydifferential
operators truncate to the derivations $\partial_t$, $|z|~\partial$,
$|z|~\overline{\partial}$, $\partial_\zeta$. The semiclassical metric
$h$ induced by the noncommutative frame is thus
\beq
\mbox{$\frac12$}\,h_{ij}(x)~\dd x^i~\dd x^j~=~-\dd t^2+|z|^{-2}~
|\dd z|^2+\dd\zeta^2~=~-\dd t^2+(\dd\log r)^2+\dd\phi^2+\dd\zeta^2
\label{classmetricframe}\eeq
with $z=r~\e^{\ii\phi}$. The semiclassical geometry of the D3-brane is
thus $\real^{1,2}\times\S^1$, consistent with the Melvin twist
construction, with a singularity at $r=|z|=0$. The measure $\Omega$
may thus be taken to be the corresponding riemannian volume form
\beq
\Omega~=~\sqrt{\det(h)}~\dd^4x~=~\mbox{$\frac1r$}~\dd t\wedge\dd r
\wedge\dd\phi\wedge\dd\zeta \ .
\label{OmegaMelvin}\eeq
The pole in the geometry at $r=|z|=0$ arises from the second
differential equation in (\ref{divtheta0Melvin}) and is unavoidable
for any cyclic measure $\Omega$~\cite{Moller1}. It is due to the
degeneracy of the Seiberg-Witten bivector field $\theta$ in
(\ref{GthetaMelvin}) at $\mx={\mbf 0}$.

Melvin backgrounds in string theory are generically unstable and can
decay via the nucleation of monopole-antimonopole pairs. This
instability may be attributed to the breaking of translation
invariance by the non-constant noncommutativity, which is incompatible
with the supersymmetry algebra. There are analogs of Prasad-Sommerfeld
monopoles in this gauge theory whose sizes scale with the
noncommutativity bivector field $\theta$ and are therefore position
dependent~\cite{HashThom2}. This feature, among others, reflects the
fate of the worldvolume theory of D-branes in the decaying Melvin background as
a noncommutative gauge theory with explicit time dependence. The only
remnant of this time dependence appears to be in the poles in the
metric induced by the noncommutative frame. In the dual supergravity
theory, such singularities manifest themselves as a discontinuity in
the open string metric along the light-cone
direction~\cite{HashThom2}.

Most known examples of non-constant Seiberg-Witten bivector fields in
string theory can be realized as a Melvin twist of a flat
D-brane~\cite{HashThom1}. In
fact, even Moyal spaces arise from a Melvin ``shift'' twist. We will
consider a further example in Section~\ref{NWpp} below. The
corresponding noncommutative field theories built on these spaces
generically exhibit violations of energy and momentum conservation
which become intertwined with quantum effects, such as UV/IR mixing,
in an intriguing way~\cite{Bert1,Robbins1}.

\subsection{Nappi-Witten Plane Wave\label{NWpp}}

The example of the Melvin universe, while providing a nice model of
non-constant noncommutativity, hides the interesting time-dependence of
the background which appears to be smoothened out in the decoupling limit
of the worldvolume gauge theory. We will now consider an example which
explicitly exhibits a time-dependent noncommutativity and can thereby
unveil interesting new physics. The system in question is the
worldvolume theory of D-branes in the Nappi-Witten background~\cite{NW1}, which
can be viewed as a monochromatic plane wave in four dimensions
supported by a null background NS--NS three-form flux $H$. One
interest in this model is that it is an {\it exact} background of
string theory, i.e. the worldsheet beta-functions
vanish to all orders, while at the same time providing a curved space
with the signature of four-dimensional Minkowski spacetime. It can
also be generated by combining the chain of dualities leading to the
Melvin universe (\ref{Melvinuniverse}) with a
boost~\cite{HashThom1}. Qualitatively, one finds a similar
noncommutative space to that generated by the Melvin universe. In
polar coordinates for the transverse space to the plane wave one finds
constant noncommutativity, while in cartesian
coordinates one obtains a Poisson representation of the
$\mathfrak{iso}(2)$ Lie algebra which is non-constant but
time-independent. Only in this situation does the background admit the
trivial gauge field configuration with curvature $F=0$ as a consistent
solution to the Born-Infeld equations of motion on the
D3-brane~\cite{HashThom1}. We will now describe a regime of the open
string dynamics in which a time-dependent noncommutativity
appears~\cite{DN1,HallSz1}, while still providing a consistent
background of string theory.

The Nappi-Witten spacetime may be defined as the group manifold of the
universal central extension of the euclidean group ${\rm ISO}(2)$ in two
dimensions~\cite{NW1}. Its non-semisimple Lie algebra is generated by elements
$J$, $T$, $P_\pm$ subject to the non-vanishing commutation relations
\beq
[P_+,P_-]~=~2\ii T \qquad \mbox{and} \qquad [J,P_\pm]~=~
\pm\ii P_\pm \ .
\label{NWcommrels}\eeq
This is just the three-dimensional Heisenberg algebra extended by an
outer automorphism which may thought of as the Fock space number
operator. These Lie brackets define a solvable algebra which we denote
by $\mfnw(4)$. The corresponding simply connected Lie group is denoted
$\NW(4)$.

Up to a Lie algebra automorphism there is a unique, non-degenerate
inner product on $\mfnw(4)$ of Minkowski signature, which can be used
to endow the group manifold of $\NW(4)$ with a homogeneous,
bi-invariant lorentzian metric. This gives the Nappi-Witten spacetime
the structure of a Cahen-Wallach symmetric spacetime in four
dimensions, whose plane wave metric in Brinkman coordinates reads
\beq
\dd s^2=2~\dd x^+~\dd x^-+|\dd z|^2-\mbox{$\frac14$}\,\vartheta^2\,
|z|^2\,\big(\dd x^+\big)^2
\label{NWBrinkman}\eeq
where $x^\pm\in\real$ parametrize the wavefront and
$(z,\overline{z}\,)\in\complex$ are coordinates on the transverse
plane. The spacetime is further supported by a $B$-field of constant
field strength
\beq
H~=~2\ii\vartheta~\dd x^+\wedge\dd z\wedge\dd\overline{z}~=~\dd B
\qquad \mbox{with} \qquad B~=~2\ii\vartheta\,x^+~
\dd z\wedge\dd\overline{z}
\ ,
\label{NWBfield}\eeq
defined to be non-vanishing only on vector fields tangent to conjugacy
classes of the group $\NW(4)$. Let us now introduce the one-form
\beq
\Lambda:=-\ii\big(\vartheta^{-1}\,x_0^-+\vartheta\,x^+\big)\,
\big(z~\dd\overline{z}-\overline{z}~\dd z\big)
\label{LamdbaNWdef}\eeq
on the null hypersurfaces of constant light-cone position $x^-=x_0^-$,
and compute the corresponding two-form gauge transformation of the
$B$-field in (\ref{NWBfield}) to get
\beq
B~\longmapsto~B+\dd\Lambda=-\ii\vartheta~\dd x^+\wedge\big(
z~\dd\overline{z}-\overline{z}~\dd z\big)+2\ii\vartheta\,x_0^-~
\dd\overline{z}\wedge\dd z \ .
\label{NWBgaugetransf}\eeq
Applying the Seiberg-Witten formula (\ref{GthetagB}) to the closed
string background fields (\ref{NWBrinkman}) and (\ref{NWBgaugetransf})
we compute~\cite{HallSz2}
\bea
\mbox{$\frac12$}\,G_{ij}(x)~\dd x^i~\dd x^j
&=&2~\dd x^+~\dd x^-+\mbox{$\frac{\vartheta^2+(x_0^-)^2}
{\vartheta^2}$}~|\dd z|^2+2\ii x_0^-\,\big(z~\dd\overline{z}-
\overline{z}~\dd z\big)~\dd x^+ \ , \nonumber\\
\theta&=&-\mbox{$\frac{2\ii\vartheta}{\vartheta^2+(x_0^-)^2}$}\,\Bigl[
\vartheta^2~\partial_-\wedge\big(z~\partial-\overline{z}~
\overline{\partial}\,\big)+4x_0^-~\partial\wedge\overline{\partial}\,
\Bigr] \ ,
\label{NWSWfields}\eea
with $\partial_\pm:=\frac\partial{\partial x^\pm}$.

For $x_0^-=0$ we recover the geometry obtained from the null Melvin
twist, with flat open string metric $G$ on the D3-brane. At the
special value $x_0^-=\vartheta$ and with the rescaling
$z\to\sqrt{2/\vartheta\,\tau}~z$, the metric $G$ in (\ref{NWSWfields})
becomes that of $\NW(4)$ in global coordinates~\cite{HallSz2} while
the non-vanishing Poisson brackets corresponding to the bivector field
$\theta$ read
\beq
\{z,\,\overline{z}\,\}_\theta~=~2\ii\vartheta~\tau \ , \quad
\{x^-,z\}_\theta~=~-\ii\vartheta~z \qquad \mbox{and} \qquad
\{x^-,\,\overline{z}\,\}_\theta~=~\ii\vartheta~\overline{z} \ .
\label{NWPoissonalg}\eeq
The Poisson algebra thereby coincides with the Nappi-Witten Lie
algebra $\mfnw(4)$ in this case and the metric on the brane with the
standard curved geometry of the pp-wave. In the semi-classical flat
space limit $\vartheta\to0$ describing the topological regime of the
open string dynamics, the quantization of $\NW(4)$ is given by the
associative Kontsevich star-product in the guise of the Gutt product
on the dual $\mfnw(4)^\vee$. With a slight abuse of notation, let us
denote the central coordinate $\tau$ of the Poisson algebra
(\ref{NWPoissonalg}) as the light-cone time coordinate $x^+$. The
semi-classical quantization is then valid in the small time limit
$x^+\to0$.

The noncommutative geometry thus obtained is an extension of that
of the Melvin universe constructed in Section~\ref{MelvinTwist} above
by explicit time dependent terms, associated to the central extension
of the Lie algebra $\mathfrak{iso}(2)$ by the generator $T$. The
Weyl-ordered Gutt product in this case turns out to be rather
complicated~\cite{HallSz1}. The generic qualitative features are best
captured by the natural ``time-symmetric'' ordering which is a
modification of the ordering used in (\ref{twistMelvin}) and is
defined by symmetrizing any monomial in $U(\mfnw(4))$ over the two
orderings obtained by placing the time translation generator $J$ to
the far right and to the far left. This is the ordering that leads to
the Brinkman form (\ref{NWBrinkman}) of the plane wave
metric~\cite{HallSz1}. The corresponding group products are worked out
exactly as described in Section~\ref{MelvinTwist} above, with the
central element $T$ generating an abstract one-parameter subgroup
acting as $\e^{\ii
  t\,T}(z,\,\overline{z}\,)=\e^{-\vartheta\,t}\,(z,\,\overline{z}\,)$
on $(z,\,\overline{z}\,)\in\complex$. The corresponding Gutt product
$*$ is again a twisted product, this time determined by the twist
element~\cite{HallSz1}
\bea
\mcF_*&=&\exp\left\{\ii\vartheta\,x^+\,\left(\e^{-\frac{\ii\vartheta}2
\,\partial_-}\,\partial\otimes\e^{-\frac{\ii\vartheta}2\,\partial_-}\,
\overline{\mdell}-\e^{\frac{\ii\vartheta}2\,\partial_-}\,
\overline{\mdell}\otimes\e^{\frac{\ii\vartheta}2\,\partial_-}\,
\mdell\right)\right.\nonumber
\\ &&\qquad\qquad+\,\,\overline{z}\,\left[\partial\otimes
\left(\e^{-\frac{\ii\vartheta}2\,\partial_-}-1\right)
+\left(\e^{\frac{\ii\vartheta}2\,\partial_-}-1\right)\otimes
\partial\right]\nonumber\\ &&\qquad\qquad+\left.
z\,\left[\,\overline{\partial}\otimes
\left(\e^{\frac{\ii\vartheta}2\,\partial_-}-1\right)
+\left(\e^{-\frac{\ii\vartheta}2\,\partial_-}-1\right)
\otimes\overline{\partial}\,\right]\right\} \ .
\label{twistNW}\eea
As before one can explicitly construct the invertible differential operator
$\mcD_*$ which implements the cohomological equivalence between the
star-products $*$ and $\star$ as asserted by the formality theorem.

A global pseudo-orthonormal frame is provided by the vector fields
\beq
e_-~=~\mbox{$\frac\partial{\partial x^-}$} \ , \quad
e_+~=~\mbox{$\frac\partial{\partial x^+}+\frac18\,\vartheta^2\,
|z|^2~\frac\partial{\partial x^-}$} \ , \quad e~=~\mbox{$\frac\partial
{\partial z}$} \qquad \mbox{and} \qquad \overline{e}~=~
\mbox{$\frac\partial{\partial\overline{z}}$} \ .
\label{orthoframeNW}\eeq
However, the construction of a compatible noncommutative
frame is much more involved than for the Melvin universe. Some
insight can be gained by examining the spacetime symmetries of the
noncommutative plane wave, which are far richer than those of the
Melvin geometry since the present background arises from a Lie group
with a bi-invariant metric. Classically, the isometry group of the
Nappi-Witten gravitational wave is the group
$\NW(4)\times\overline{\NW(4)}$ induced by the left and right regular
actions of the Lie group $\NW(4)$ on itself. The corresponding Killing
vector fields live in the seven-dimensional Lie algebra
$\mfg:=\mfnw\oplus\overline{\mfnw}$. The left and right actions
of the central element $T=\vartheta\,\partial_-$ coincide and generate
translations in the light-cone position $x^-$. The Killing vector
field ${J}=\vartheta^{-1}\,\partial_+$ generates time
translations along $x^+$, while $J+\overline{J}=
-\ii(z\,\partial-\overline{z}\,\overline{\partial}\,)$ generates rotations
in the transverse $(z,\,\overline{z}\,)$-plane. The remaining four
vector fields generate ``twisted'' translations in the transverse
plane~\cite{HallSz2} which are completely analogous to the twisted
translational symmetry of a planar system subject to a constant,
perpendicularly applied magnetic field of strength
$\vartheta\,x^+$~\cite{Szrev2}. Remarkably, Lorentz boosts are not
amongst these symmetries even in the flat space limit
$\vartheta\to0$.

Let us now describe the corresponding twisted isometries. For brevity
we will only consider translation generators. Using the twist element
(\ref{twistNW}) we arrive at the twisted coproducts~\cite{HallSz1}
\bea
\Delta_*(\partial_-)&=&\partial_-
\otimes1+1\otimes\partial_- \ , \nonumber\\
\Delta_*(\partial_+)&=&\partial_+\otimes1+
1\otimes\partial_++\vartheta\,\bigl(\e^{-\frac{\ii\vartheta}2\,
\partial_-}\,\mdell\otimes
\e^{-\frac{\ii\vartheta}2\,\partial_-}\,\overline{\mdell}-
\e^{\frac{\ii\vartheta}2\,\partial_-}\,
\overline{\mdell}\otimes\e^{\frac{\ii\vartheta}2\,
\partial_-}\,\mdell\bigr) \ , \nonumber\\
\Delta_*(\partial)&=&\partial\otimes
\e^{-\frac{\ii\vartheta}2\,\partial_-}+\e^{\frac{\ii\vartheta}2\,
\partial_-}\otimes\partial \ , \nonumber\\
\Delta_*\big(\,\overline{\partial}\,\big)&=&
\overline{\partial}\otimes
\e^{\frac{\ii\vartheta}2\,\partial_-}+\e^{-\frac{\ii\vartheta}2\,
\partial_-}\otimes\overline{\partial} \ .
\label{STOcoprods}\eea
An action of the spacetime translations which is compatible with the
noncommutative algebra of functions on $\NW(4)$ is given by
\bea
\partial_-^*\triangleright f&=&\partial_-f \ , \nonumber\\
\partial_+^*\triangleright f&=&\partial_+f \ , \nonumber\\
\partial^*\triangleright f&=&\e^{-\frac{\ii\vartheta}2\,
\partial_-}\,\partial f \ , \nonumber\\
\overline{\partial}\,{}^*\triangleright f&=&
\e^{\frac{\ii\vartheta}2\,\partial_-}\,
\overline{\partial}f \ .
\label{STOderivs}\eea
From (\ref{STOcoprods}) we see the breaking of the classical time
translation invariance by the time-dependent NS--NS background
(\ref{NWBfield}), while (\ref{STOderivs}) further exhibits the
twisting of the transverse plane translations by the magnetic field
$\vartheta\,x^+$. On the other hand, the classical translational
symmetry of the spacetime along the light-cone position persists in
the quantum geometry.

One particularly noteworthy aspect of the construction of
noncommutative gauge theory on the present spacetime concerns the
possible choices of integration measure $\Omega$. In this case one can
reduce the divergence-free conditions (\ref{divtheta0}) to the
equations~\cite{HallSz1}
\beq
\partial_-\Omega~=~0 \qquad \mbox{and} \qquad z\,\partial\Omega~=~
\overline{z}\,\overline{\partial}\Omega \ .
\label{NWmeascond}\eeq
In contrast to (\ref{divtheta0Melvin}), it is possible to find
non-singular solutions to the differential equations
(\ref{NWmeascond}). Consistency between differential operator and
function star-commutators demands that $\Omega$ be a function of the
light-cone time $x^+$ alone. In particular, the flat choice
$\Omega=\dd^4x$ is possible. This is rather remarkable, in that it
provides an example of a Lie algebra which is {\it not} nilpotent, yet
for which the cyclicity property (\ref{Kontcycl2}) holds for the Gutt
product $*$ without any modification of the flat space measure. In
this case the enhanced isometry group of the plane wave, arising from
the central extension, ``flattens'' out the singularities of geometries
like the one of Section~\ref{MelvinTwist} above.

The noncommutative gauge theory that we have thus far described is the
worldvolume theory of a non-symmetric curved D3-brane wrapping all of
the $\NW(4)$ spacetime. Let us now describe the noncommutative gauge
theory on regularly embedded worldvolumes of D-branes in $\NW(4)$. The
branes of interest are the spacelike D-strings (or {\it S1-branes})
which wrap untwisted conjugacy classes of the Nappi-Witten
group~\cite{F-OFS1}. The noncommutative gauge theory of these branes
can be obtained by using the general formalism of
Section~\ref{ABranes} to restrict the geometry of $\NW(4)$ above to
obtain the usual quantization of coadjoint orbits in
$\mfnw(4)^\vee$~\cite{HallSz2}. In exactly the same way that the
noncommutative space $\real_\infty^3$ of Section~\ref{S3DBranes} above
can be viewed as a collection of all fuzzy spheres, we can regard the
noncommutative geometry of $\NW(4)$ as a foliation by all
noncommutative S1-branes.

The non-degenerate conjugacy classes of the group $\NW(4)$ are
coordinatized by the transverse plane
$(z,\,\overline{z}\,)\in\complex\cong\real^2$. They are defined by the
spacelike planes of constant time in $\NW(4)$ given by the transversal
intersections of the null volumes~\cite{F-OFS1}
\beq
x^+~=~{\rm constant} \qquad \mbox{and} \qquad
x^-+\mbox{$\frac14$}\,\vartheta\,|z|^2\,\cot\bigl(\mbox{$\frac12$}\,
\vartheta\,x^+\bigr)~=~{\rm constant} \ .
\label{S1constrs}\eeq
This describes the brane worldvolume as a wavefront expanding in a
circular Larmor orbit in the transverse plane. We will find below in
fact that the S1-branes are completely analogous to branes in flat
space with a magnetic field on their worldvolume~\cite{Szrev2}. In the
semiclassical limit $\vartheta\to0$, the second constraint in
(\ref{S1constrs}) to leading order becomes
\beq
C~:=~2\,x^+\,x^-+|z|^2~=~{\rm constant} \ .
\label{Cconstr}\eeq
The function $C$ corresponds to the quadratic Casimir element of
$U(\mfnw(4))$ and the constraint (\ref{Cconstr}) is analogous to the
requirement that Casimir operators act as scalars in irreducible
representations. Similarly, the constraint on the time coordinate
$x^+$ in (\ref{S1constrs}) is analogous to the requirement that the
central element $T$ act as a scalar operator in any irreducible
representation of $\NW(4)$.

To apply the noncommutative version of Poisson reduction described in
Section~\ref{ABranes}, we first need project the algebra of functions
onto the star-subalgebra of functions which star-commute with the
Casimir function $C$. These are naturally the fields $f$ which are
independent of the light-cone position so that
$\partial_-f=0$~\cite{HallSz1}. One then finds that the star-product
$*$ determined by the twist element (\ref{twistNW}) restricts to the
{\it Moyal} product $\star_0$, with noncommutativity parameter
$\vartheta\,x^+$, on the S1-branes. This is expected from the form of
the restricted open string fields (\ref{NWSWfields}) in this
case. Using (\ref{redsymmetries}), (\ref{STOcoprods})
and (\ref{STOderivs}) one recovers the expected actions of
translations $\partial^{\star_0}\triangleright f=\partial f$ and
$\overline{\partial}\,{}^{\star_0}\triangleright
f=\overline{\partial}f$ on the Moyal plane, with primitive coproducts
$\Delta_{\star_0}$ appropriate to the translational symmetry of
canonical noncommutative field theory. Consistent with the reduction
to the conjugacy classes, one also finds $\partial_\pm^{\star_0}\triangleright
f=0$. What is particularly interesting about the reduction from $\NW(4)$
is that one arrives at a non-vanishing co-action of time translations
given by
\beq
\Delta_{\star_0}(\partial_+)=\vartheta\,\big(\partial\otimes
\overline{\partial}-\overline{\partial}\otimes\partial\big) \ .
\label{delplusnon0}\eeq
Recalling that $J=\vartheta^{-1}\,\partial_+$ and that the vector
field $J+\overline{J}$ generates rotations of the transverse plane, we
see that the time translation isometry of $\NW(4)$ truncates to
rotations and (\ref{delplusnon0}) is just the standard twisted
coproduct of rotations of the Moyal plane that we encountered in
Section~\ref{TwistPoincare}. Thus the embedding of standard
noncommutative field theories into gravitational waves naturally
endows them with twisted spacetime symmetries.

\subsection*{Acknowledgments}

The author is grateful to T.R.~Govindarajan, G.~Landi, F.~Lizzi,
M.~Riccardi and C.~S\"amann for helpful discussions. This work was
supported in part by PPARC Grant PPA/G/S/2002/00478 and by the EU-RTN
Network Grant MRTN-CT-2004-005104.

\end{document}